\documentclass[11pt,a4paper]{JHEP3}
\pdfoutput=1
\usepackage{graphicx}
\usepackage{bm}
\usepackage{mathrsfs}
\usepackage{amsmath}
\usepackage{amssymb}
\usepackage{amstext}
\usepackage{dsfont}
\usepackage{epsfig,latexsym}

\newcommand{\tlrho}{{\tilde{\rho}}}

\newcommand{\tlmu}{{\tilde{\mu}}}
\newcommand{\tlt}{{\tilde{t}}} 
\newcommand{\tx}{{\tilde{x}}}

\newcommand{\baT}{{\bar{T}}}
\newcommand{\baF}{{\bar{F}}}
\newcommand{\bazet}{{\bar{\zeta}}}

\newcommand{\order}{{\mathcal{O}}}
\newcommand{\str}{\mathrm{STr}}
\newcommand{\tr}{\mathrm{Tr}}
\newcommand{\cd}{\mathcal{D}}
\newcommand{\s}{\hat{S}}

\newcommand{\varphiz}{{\varphiz^{\! z ~}}}

\newcommand{\labell}[1]{\label{#1}}
\newcommand{\reef}[1]{(\ref{#1})}
\newcommand{\dreef}[2]{(\ref{#1},\ref{#2})}
\newcommand{\myfigurer}[3]{\begin{figure}[h!]
\parbox[b]{0.5\textwidth}{\includegraphics[width = 0.5\textwidth]{#1}}\hspace{0.01\textwidth}
\parbox[b]{0.47\textwidth}{\caption{#2}\label{#3}}\vfill\end{figure}}
\newcommand{\myfigurerw}[5]{\begin{figure}[h!]
\parbox[b]{#1}{\includegraphics[width = #1]{#3}}\hspace{0.01\textwidth}
\parbox[b]{#2}{\caption{#4}\label{#5}}\end{figure}}
\newcommand{\myfigure}[3]{\begin{floatingfigure}[h!tb]\includegraphics[width = 0.49\textwidth]{#1}\caption{#2}\label{#3}\end{floatingfigure}}
\newcommand{\myfigurew}[4]{\begin{figure}[h!tb]\includegraphics[#1]{#2}\caption{#3}\label{#4}\end{figure}}
\newcommand{\myfigured}[4]{\begin{figure}[ht]\includegraphics[width = 0.48\textwidth]{#1}\hspace{0.02\textwidth}\includegraphics[width = 0.48\textwidth]{#2}\caption{#3}\label{#4}\end{figure}}
\newcommand{\myfiguredd}[6]{\begin{figure}[ht]
\begin{minipage}[t]{0.47\textwidth}
\centering
\includegraphics[width = 0.99\textwidth]{#1}\caption{#2}\label{#3}
\end{minipage}
\hspace{0.043\textwidth}
\begin{minipage}[t]{0.47\textwidth}
\centering
\includegraphics[width = 0.99\textwidth]{#4}\caption{#5}\label{#6}
\end{minipage}
\end{figure}}
\title{Three Roads to Probe-Brane Superconductivity}
\author{Bum-Hoon Lee and Matthias C. Wapler\\
Center for Quantum Spacetime, Sogang University, Seoul, Korea \\
\\E-mail: \email{bhl@sogang.ac.kr}, \email{wapler@sogang.ac.kr}}
\preprint{}
\date{\today}
\abstract{We study a defect system of two parallel D5 probe branes in a large-$N_c$ D3 background. Using the non-abelian DBI action, we study three different fields that can give rise to a superconducting phase transition: A vector (p-wave), a scalar corresponding to a non-trivial ``separation'' of the branes in the (3+1) field theory directions and a scalar corresponding to a separation in the ``internal'' $S^5$ (both s-wave).

Comparing these phases first in the $\alpha'^2$ expansion, we find that the internal scalar has the largest critical temperature and is always thermodynamically preferred. Further, there is an interesting attractor behavior.

Taking a simplified version of the full DBI action that preserves its regularity and geometry, we find that the divergences of the $\alpha'^2$ expansion are resolved and some second order transitions turn into first order ones. In addition to some other changes of the phase diagram due to the structure of the DBI action, we observe
that the ground state degeneracy of the unbroken theory is lifted.
We also isolate the unphysical artifacts of our simplification.}
\begin{document}
\tableofcontents
\section{Introduction}\label{introduction}
In recent years, the ADS/CFT correspondence \cite{adscft_Maldacena,adscft_gubser,adscft_witten} has become a powerful tool to study various properties of the strong coupling limit of conformal field theories, with applications to QCD and also more recently to some aspects of condensed matter physics. In principle, one has to differentiate between top-down setups that are constructed within string theory and imply consistency -- and bottom-up setups, where the gravitational duals are constructed from a phenomenological point of view. In this paper, we use the former approach as we would like to explore what happens to a particular consistent theory. 

There are two ways to motivate the present work:
In condensed matter applications, there has been particular interest and a significant amount of activity in 2+1 dimensional systems that can be typically constructed using M2 branes \cite{pavel}, or as a defect in a 3+1 dimensional background using D3-D5 \cite{lisa,hirosi,fancycon,conpaper,EvansBKT,KarchBKT} and D3-D7 \cite{fancycon,conpaper,rey,bergman,deogki,kraus} brane intersections -- and also as bottom-up setups in various contexts. As our world is 3+1 dimensional, the defect setup may be more realistic, even though there are some difficulties involving the stability of the D3-D7 systems \cite{conpaper,fancycon,bergman}. These are based on a common way how to introduce fundamental matter in ADS/CFT: probe brane configurations, where one considers a planar $AdS$ blackhole background. For example one considers the well-known $AdS_5\times S^5$ (planar black hole) solution (above the deconfinement phase transition) from the decoupling limit of a stack of $N_c\gg 1$ D3 branes, that is dual to a (thermal) $SU(N_c)$ $\mathcal{N}=4$ supersymmetric Yang-Mills theory \cite{adscft_gubser}. Then one inserts $N_f \ll N_c$ intersecting Dp branes, giving $N_f$ families of (charged) fields in the fundamental representation of the $SU(N_c)$, living along the directions of the intersection \cite{winters,robfirst} -- such as the above-mentioned D3-Dp intersections.

One interesting question to ask then is what happens if one takes two (or more) D5 probe branes of the (2+1) defect system and studies new effects that appear if they are physically separated in the third (3+1) direction. This is the question that evolved into the present work.

%

Another motivation are the constructions of gravitational configurations whose duals show spontaneous second-order symmetry breaking phase transitions and other properties reminiscent of superconductors \cite{gubsersmall,gubserpre,hartnollsmall,hartnollbig}. This sparked great interest and there has been a significant amount of activity in this field studying effects such as the Meissner effect and magnetic vortices (e.g. \cite{magsuperjohnson1,magsuperjohnson2,magsupertaiwan1}), Fermi effects (e.g. \cite{andyfermi,gubserfermi,johannascon5}) or higher order/backreaction effects (e.g. \cite{Johannaback,superhigher}).

Most of these constructions were however bottom-up models. While there has been also some activity on constructing duals of superconductors in supergravity theories that can be obtained from M theory \cite{imperialsuper,nakwoo,gubserstring,gubserM} with many interesting results, e.g. the ``landscape of superconductors'' \cite{denefhartnoll}. In contrast, there has been only very limited activity -- which gave rise so far only to p-wave superconductors \cite{johannascon7small,johannascon7big,UBCscon,johannascon5} -- on constructing D-brane configurations that give rise to superconductivity more directly from string theory, give very precise information on the details of the field theory dual and give always regular solutions.

In this paper, we consider two parallel defect probe D5 branes in a D3 background. In the field theory side, they correspond to a 2+1 dimensional system of two species of particles, e.g. a multi-layered system as in a cuprate superconductor or multi-layered graphene. One way to get some interesting physics from their separation in the flat directions is then to turn on a non-trivial $SU(2)$ potential in the world-volume theory. This corresponds e.g. to an isospin chemical potential/density \cite{isospin} or any other non-trivial combination of the densities of the different species. Then, we can study the phase diagram and look e.g. for a spontaneous symmetry breaking phase transition. In addition to the separation in the (3+1) physical directions, we will also consider a separation in the ``internal'' space and an $SU(2)$ magnetic field, such that we can compare the different phases and have for the first time a system in which different s-wave and p-wave superconducting condensates compete with each other. 

We will consider both the quadratic expansion of the world-volume action and an adaption of the full action that allows us to study a slightly modified version of the full non-perturbative action -- giving us e.g. consistent regular physics in the low-temperature limit. We will see how the non-perturbative effects change the phase diagram, and isolate and discuss an artifact that comes from our simplification.

The paper is organized as follows: In section \ref{setup}, we review and describe the details of the string theory configuration with its field theory dual. In section \ref{pert}, we obtain and discuss the results of the different phases in the perturbative treatment of the DBI action. In section \ref{fullact}, we describe a small modification to the action that allows us to study non-perturbative effects, apply this to the different phases and discuss the changes compared to the perturbative case. Finally in section \ref{conclusions}, we conclude and discuss the results and give some outlook for possible future extensions of this work. For completeness, we present some results related to the grand canonical ensemble in appendix \ref{gcapp}.
\section{Setup}\label{setup}
We start with the supergravity background of a planar black hole in $AdS_5$, 
\begin{eqnarray}
ds^2 & =& \frac{r^2}{L^2} \left( -(1-r_0^4/r^4)dt^2 +d\vec{x}_3^2\right)
+ \frac{L^2}{r^2} \left( \frac{dr^2}{1-r_0^4/r^4} +r^2 d\Omega_5^2\right) \
, \ \ \  C^{(4)}_{txyz}=-\frac{r^4}{L^4} \labell{D3geom} \ .
\end{eqnarray}
This
corresponds to the decoupling limit of $N_c$ black D3-branes
dual to ${\cal N}=4$ $SU(N_c)$ super-Yang-Mills theory at finite temperature $T$, living along the flat directions of the AdS spacetime \cite{adscft_witten,adscft_gubser}.
The  temperature is given by the Hawking temperature $T = \frac{r_0}{\pi L^2}$ and the Yang-Mills coupling is related to the string coupling $g_s$  by $g_{ym}^2 = 4 \pi g_s$. Since the AdS length $L$ is
given in terms of the string coupling and 
string length  $l_s$ as $L^4 = 4\pi\, g_s N_c \, l_s^4$, the 't Hooft
coupling $\lambda=g_{ym}^2 N_c$ can be written as $\lambda=\frac{L^4}{l_s^4}$. Hence the ``supergravity limit'' $L \gg l_s$ in which the type IIB supergravity action and the solution \reef{D3geom} are valid  corresponds to strong coupling $\lambda \gg 1$.

In practice, however, we will use coordinates that are made dimensionless with factors of $\frac{L^2}{r_0} =\frac{1}{\pi T}$, denoted by $\tilde{(\cdot)}$, such as $\tlt: = \frac{r_0 \, t}{L^2} =  t\, (\pi T)$, and the inverse dimensionless radius $u := \frac{r_0}{r}$, giving us 
\begin{equation}\label{branemetric}
ds^2 \, = \, \frac{L^2}{u^2}\left(-(1-u^4)d\tlt^2 + d\vec{\tx}_3^2 + \frac{du^2}{1-u^4} + u^2 d\Omega_5^2 \right) \ .
\end{equation}

In this setup all the fields transform in the adjoint representation of the $SU(N_c)$.
However in QCD or condensed matter physics, one also needs to consider matter that is charged under this symmetry, i.e. that transforms in the fundamental representation.
To introduce the fundamental matter one then creates an intersection of ``probe'' Dp branes with the D3 branes, such that in the string theory side there are massless fields in the effective field theory at the intersection. From the point of view of the probe branes, they correspond to endpoints of D3-Dp strings, and in the gravity side they correspond to fundamental fields in the (defect) field theory \cite{KarchKatz,Maldacenaprobe}.

Here, we use the well- known D3-D5 defect setup (see e.g. \cite{lisa,hirosi,johannadef}):
\begin{equation}
\begin{array}{rccccc|c|cccccl}
  & & 0 & 1 & 2 & 3 & 4& 5 & 6 & 7 & 8 & 9 &\\
  & & t & x & y & z & r&   &   &   &   &  \theta &\\
\mathrm{background\,:}& D3 & \times & \times & \times & \times & & &  & & & & \\
\mathrm{probe\,:}& D5 & \times & \times & \times &  & \times  & \times & \times & &  & &  \ \ \ .
\end{array} 
\labell{array}
\end{equation}
The dual field
theory is now the SYM gauge theory coupled to $N_f$ fundamental
hypermultiplets, which are confined to a (2+1)-dimensional defect.
This construction is still supersymmetric, but the supersymmetry has been reduced from ${\cal N}=4$ to
${\cal N}=2$ by the introduction of the defect. In the limit $N_f
\ll N_c$, the D5-branes may be treated as probes in the
supergravity background, i.e. we may ignore their gravitational
back-reaction \cite{KarchKatz,Maldacenaprobe}.

In contrast to \cite{conpaper,fancycon,fancytherm,funnel}, we want to use the non-abelian structure of the $U(N_f)$ theory for the simplest case of $N_f=2$. Furthermore, we are considering at most one scalar at one time and set the two-form $B_{\mu\nu} =0$ such that the non-abelian DBI action \cite{dielectric}
\begin{eqnarray}\labell{robbigact}
S & = & - T_p \int_{Dp} \str \sqrt{- \det(P[E+ E(Q^{-1}-\mathds{1})E ] + 2\pi l_s^2 F )\det(Q E)} \ \nonumber \\
& & ~~~~~~~~~~~~~~ + \ \mu_p \int_{Dp}\tr \left(P\left[e^{i 2\pi l_s^2 i_\Phi i_\Phi} \sum_{n}C^{(n)} \right]e^{i 2\pi l_s^2 F} \right) \ , \\
Q^{\mu\nu} & = & E^{\mu\nu} \, + \,  2\pi i [\Phi^\mu,\Phi^\nu] \ , \ \ E_{\mu \nu}\, =\, G_{\mu \nu} \, +\,  B_{\mu \nu} 
\end{eqnarray}
reduces to
\begin{equation}\label{braneaction}
S \, = \, - T_5  \int_{D5} \str \sqrt{- det(P[G] + 2\pi l_s^2 F )}   \ ,
\end{equation}
evaluated in the D3 background \reef{D3geom}.

Here, we want to consider the scalars in the $z$ direction (i.e. along the D3 branes), $\Phi^z$, and the spherical direction $\Phi^{\sin\theta}$ (away from the D3s) that describe the position and separation of the probe branes. The Pauli matrices are (with an additional normalization factor):
\begin{equation}\labell{paulidef}
\tau_0 = \frac{1}{2}\left\{ \begin{array}{c c} 1 & 0 \\ 0 & 1\end{array} \right\}\ , \ \ 
\tau_1 = \frac{1}{2}\left\{ \begin{array}{c c} 0 & 1 \\ 1 & 0\end{array} \right\}\ , \ \ 
\tau_2 = \frac{1}{2}\left\{ \begin{array}{c c} 0 & - i \\ i & 0\end{array} \right\}\ , \ \ 
\tau_3 = \frac{1}{2}\left\{ \begin{array}{c c} 1 & 0 \\ 0 & -1\end{array} \right\} \ ,
\end{equation}
with the properties
\begin{equation}\labell{su2rel}
[\tau_a,\tau_b] = i \epsilon_{ab}^c \tau_c \ , \ \ \{\tau_a,\tau_b \}  = \delta_{ab} \tau_0 \, \ \ [\tau_a,\tau_0] = 0\ \mathrm{and} \ \ \{\tau_a,\tau_0 \} =  \tau_a \ .
\end{equation}
In particular, we want to consider configurations that are isotropic in the defect directions and static, such that the only coordinate dependence is on $u$ and we also we want to turn on only one of the non-trivial $SU(2)$ generators
and as our results will be invariant under an overall $SU(2)$ rotation, we can choose e.g. $\tau_1$. Choosing the parametrization of the $S^5$ sphere as $d\Omega_5^2 \, = \, d\theta^2 \, + \, \sin^2 \theta d\Omega_2^2\, + \, \cos^2 \theta d\Omega_2^2$ and putting the D5 on the second 2-sphere, we will choose the Ansatz
\begin{equation}\label{scalaransatz}
\Phi^z \, = \, 2 \psi(u) \tau_1 \ \mathrm{and} \ \ \ \Phi^{\sin\theta} \, = \, 2 \phi(u) \tau_1 \ .
\end{equation}
This gives the induced metric (if we fallaciously evaluate the multiplication of the $SU(2)$ generators already now)
\begin{equation}
ds^2 \, = \, \frac{L^2}{ u^2} 2 \tau_0 \left(-(1-u^4)d\tlt^2 + d\vec{\tx}_2^2 + \left(\frac{1}{1-u^4} +\psi'^2 +\frac{u^2\phi'^2}{1-\phi^2}   \right)du^2  \right) \, + \, L^2 (1-\phi^2) d\Omega_2^2 \ ,
\end{equation}
where $(\cdot)'$ denotes $\partial_u (\cdot)$.
Just as a motivation let us look at the case of the $\Phi^z$ scalar only. The solution is just
\begin{equation}
\partial_u \psi \ =\   \frac{ c }{\sqrt{(1-u^4)((1-u^4) - c^2 u^8)}}
\end{equation}
for some constant $c$ and some $U(2)$ valued matrix $\tau$. This tells us that without additional fields, the only embedding in which the branes fall into the horizon is the trivial one. In the only other case, the branes extend out to $\psi = \pm \infty$  at $u^4 = \frac{- 1}{2 c^2} +\sqrt{\frac{1}{4 c^4}+\frac{1}{ c^2}} $. In order to find non-trivial embeddings, we should then turn on also a gauge field that does not commute with the scalar.

To get a simple physical interpretation, we choose the gauge field to be proportional to $\tau_3$ with the Ansatz
\begin{equation}\labell{elansatz}
A \, = \, \frac{L^2}{\pi l_s^2} \rho(u) \tau_3 d\tlt \ ,
\end{equation}
which can be interpreted as the gravity dual of the isospin density \cite{isospin}. As the isospin number is essentially given by the appropriate number of string end/``start'' points on the probe branes, the isospin density and isospin chemical potential are then given by
\begin{equation}\label{onemoredef}
\lim_{u\rightarrow 0} \partial_u\rho(u) \, = \, \tlrho \, = \, \frac{1}{2 N_c T^2}\rho_{iso} \ \mathrm{and} \ \ \ \lim_{u\rightarrow 0} \rho(u) \, = \,\tlmu \, = \, \frac{2 }{\sqrt{\lambda} T} \mu_{iso} \ .
\end{equation}
Taking into account the newly arising commutators through the covariant derivative $\mathcal{D}_\mu  \Phi = \partial_\mu \Phi + i [A_\mu, \Phi]$, the pullback of the background metric \reef{branemetric} becomes
\begin{eqnarray}\nonumber\labell{branesubmetric}
ds^2 & = & 2 \frac{L^2}{ u^2}\left(-(1-u^4) d\tlt^2 +  d\vec{\tx}_2^2 + \frac{du^2}{1-u^4}   \right)\tau_0\, +  \\ \nonumber
& & 2 \frac{L^2 }{u^2}\left( - 2 (\psi \rho \tau_3)(\psi \rho \tau_3) dt^2 + 4 i\{\psi \rho \tau_3,\psi' \tau_1 \} dt du  + 2 (\psi' \tau_1)(\psi' \tau_1) du^2 \right)\, + \\ \nonumber
& & \frac{2 L^2}{1-\phi^2}\left( - 2 (\phi \rho \tau_3)(\phi \rho \tau_3) d\tlt^2 + 4 i\{\phi \rho \tau_3,\phi' \tau_1 \}d\tlt du  + 2 (\phi' \tau_1)(\phi' \tau_1) du^2 \right) \\ 
& & ~~~~~~~~~~~~~~~~~~~~~~~~~~ \, + \, 2 \tau_0 L^2 (1-\phi^2) d\Omega_2^2  \ ,
\end{eqnarray}
where we kept the generators explicitly because one has to take the symmetrized trace. 

To complete the picture, we will also alternatively to the scalars consider a ``magnetic'' field, taking the Ansatz
\begin{equation}\labell{vectoransatz}
A \ =\ \frac{L^2}{2 \pi l_s^2} \left(2 \rho(u) \tau_3 d\tlt\, +\, \omega(u) \tau_1 d\tx\right) \ .
\end{equation}

In order to determine the phase diagram, we obtain the free energy in the usual way from the regularized euclidean on-shell action. By studying the variation of the fields, one can trivially see that this is actually a function of the chemical potential, rather than the density, so we get \cite{bigrev,robchem,robbig,long,johannaphase,fancytherm}:
\begin{equation}
\Omega(T,\mu) \, = \, T I_e^{reg.} \ .
\end{equation}
The counter terms that we need to add are dictated to us by requiring consistency of the variational principle and give in analogy to the abelian case \cite{skenderis}
\begin{equation}\label{ibdy}
I_{bdy.} \ = \ - \frac{1}{3} \sqrt{\gamma} + \frac{1}{2}\phi^2\sqrt{\gamma} \ .
\end{equation}
As we will see later, it turns out that the other scalar $\psi$ does not contribute to the boundary term.
The Helmholtz free energy can then be obtained from a Legendre transformation, 
\begin{equation}
F(T,\rho) \, = \, \Omega \, + \, \rho \mu \ .
\end{equation}
As it seems most physical, and also for reasons that we will see later, we want to fix the isospin density and hence work in the canonical ensemble using the Helmholtz free energy. 

To obtain some dimensionless quantity that naturally appears in terms of $\tlrho$ or $\tlmu$ and the dimensionless boundary values for $\omega$, $\phi$ and $\psi$, we can define the usual dimensionless
\begin{equation}
\tilde{F} \, := \, \frac{F}{\sqrt{\lambda} N_c T^3} \ \mathrm{and} \ \ \ \tilde{\Omega} \, := \, \frac{\Omega}{\sqrt{\lambda} N_c T^3} \ .
\end{equation}
Since it is for most processes however more physical for a superconductor to fix the particle number rather than the temperature, the dimensionless quantity
\begin{equation}
\bar{F} \ = \ \frac{\tilde{F}}{\tlrho^{3/2}} \ = \ \frac{2 \sqrt{2 N_c}}{\sqrt{\lambda} \rho^{3/2}}
\end{equation}
is the more physically relevant one.
In oder to obtain a positive value, we define the energy gap between the unbroken and broken phases as
\begin{equation}
\Delta \bar{F} \ = \ \bar{F}_{unbroken} \ - \ \bar{F}_{broken} \ .
\end{equation}
In appendix \ref{gcapp}, we will also refer to the Gibbs free energy, where we define the energy gap exactly the other way round, $\Delta \tilde{\Omega} \, = \, \tilde{\Omega}_{broken}-\tilde{\Omega}_{unbroken}$.
%
%
\section{Perturbative expansion}\label{pert}
In the following, we look for convenience only at the leading $\alpha'$ expansion, where we scaled the scalars $\Phi^a$ such that they appear at the same order of $\alpha'$ as the gauge field $A$. The action for this pertubation is then
\begin{equation}\labell{appact}
S \, = \, - \frac{1}{2\pi^3}\sqrt{\lambda}N_c\int d^4 \tilde{\sigma}  \sqrt{-g}\str \left( 2 \cd_\mu \Phi^{\sin\theta} \cd^\mu \Phi^{\sin\theta}  -  4 \Phi^{\sin\theta}\Phi^{\sin\theta}  +  \frac{2}{u^4} \cd_\mu \Phi^z \,\cd^\mu \Phi^z   +  F^2  \right) \ ,
\end{equation}
where the metric $g$ is just the induced metric without the scalar perturbations. 
Substituting in the Ansatz \dreef{scalaransatz}{vectoransatz} and the metric \reef{branesubmetric}, we obtain 
\begin{equation}\labell{appactsub}
S  =  \frac{\sqrt{\lambda}N_c \tilde{V}_{2+1}}{2\pi^3} \int_0^1\!\!\! du \frac{1}{u^4}\left(\! \rho^2\frac{\psi^2\! +\! 4u^2\phi^2\! +\! u^4 \omega^2}{1-u^4}  +  8\phi^2 + u^4 \rho'^2 +  (1\! -\!\! u^4)\left(\psi'^2 \! +\! 4u^2\phi'^2\! +\! u^4 \omega'^2 \right)\! \right) \ .
\end{equation}
The volume factor $\tilde{V}_{2+1}$ for the dimensionless ``flat'' coordinates that we introduced here gives rise to a factor of $2\pi^3 T^3$ in the densitized euclidean action.

The solution in the case where we turn on only the isospin potential is trivially
\begin{equation}
\rho(u) \ = \ q (1-u) \ , \ \ \tlrho = \tlmu = q 
\end{equation}
for some number q.
This gives the free energy $\tilde{F} \, = \, - \frac{1}{2} \tlrho^2$.

Before we start using this action to obtain the phase diagram, let us review the appropriate first basic examples in the literature.
\subsection{P-Wave superconductor}\label{ppert}
The p-wave superconductor was found in a bottom-up approach in \cite{gubsersmall} and implemented in a D3-D7 system with 2 D7 probes in \cite{johannascon7small,UBCscon,johannascon7big} and in our setup in \cite{johannascon5}. Up to a difference of $1-u^3$ vs. $1-u^4$ in the blackhole factor the bottom-up and top-down models should be the same at the perturbative level.

This case is called P-wave, because the symmetry breaking is caused by a vector field and the resulting spectral curves are highly anisotropic \cite{gubsersmall}. As an Ansatz for the gauge field one can then choose $A = \rho(u) \tau_3 d\tlt + \omega(u) \tau_1 d\tx$, where $\omega(u)$ will play the role of the symmetry breaking condensate, and one could certainly choose any other combination of non-commuting generators.

In \cite{gubsersmall}, the equations of motion are (in our notation):
\begin{equation}
\left(u^2 \rho'(u) \right)' \, = \, \frac{\omega(u)^2}{1- u^3}\rho(u) \ \mathrm{and} \ \ 
\left((1-u^3) \omega'(u)\right) \, = \, \frac{\rho(u)^2}{1-u^3} \omega(u) \ .
\end{equation}
The expansions near $u=0$ are 
\begin{equation}\label{ominf}
\rho(u)  \, = \, \tlmu - \tlrho \,u + \ldots \ \mathrm{and} \ \ \omega(u) \, = \, \omega_\infty + \tilde{\zeta}_\omega \, u +\ldots
\end{equation}
and close to the horizon they are:
\begin{equation}
\rho(u) \, = \, - q (u-1) + \ldots \ \mathrm{and} \ \ \omega(u) \, = \, \omega_0 + \order(1-u)^2 \ ,
\end{equation}
the first of which arises from requiting $|A|$ to be finite on the horizon, and the second because the equation of motion reduces to first order on the horizon.
For a consistent field theory, $\omega_\infty$ must vanish. Hence we are left with a one-parameter solution that traces out a line in the $(q,\omega_0)$ plane or the $(Q,\Omega_1)$ plane. Obviously $\mu$ corresponds to a chemical potential and $Q$ to a charge density, and $\zeta_\omega \propto <J^1_x>$ makes a good candidate for the order parameter (see eq. \reef{onemoredef}). In practice we will always represent the condensate scaled by an appropriate factor of the density, in this case $\bar{\zeta}\omega \, = \, \frac{\tilde{\zeta}_\omega}{\sqrt{\tlrho}}$.

In our case, the equations of motion can be straightforwardly obtained from \reef{appact} and become
\begin{equation}
\left(u^2 \rho'(u) \right)' \, = \, \frac{\omega(u)^2}{1- u^4}\rho(u) \ \mathrm{and} \ \ 
\left((1-u^4) \omega'(u)\right) \, = \, \frac{\rho(u)^2}{1-u^4} \omega(u) \ 
\end{equation}
but the near-horizon and asymptotic expansions remain the same.
\subsection{S-Wave superconductor}\label{spert}
In the case of the s-wave superconductor, the symmetry-breaking operator is a charged scalar in a charged planar black hole background and it was first found in a bottom-up setup in \cite{hartnollsmall,hartnollbig}. Later, for example various M-theory compactifications were studied as to whether they obey superconducting behavior in \cite{denefhartnoll}.

The action in \cite{hartnollsmall} was
\begin{equation}
\mathcal{L} \, = \, - \, \frac{1}{4} F^2 + \frac{2|\Phi|^2}{L^2} \, - \, |\mathcal{D}\Phi|^2
\end{equation}
with the equations of motion
\begin{eqnarray}\labell{hartnolleom}
0 & = & \partial_u\left((u^{-2} - u) \partial_u \phi\right)\, + \, u^{-4} \left(\frac{\rho^2}{u^{-2} -u} + 2 \right) \phi \ , \\
0 & = & \partial_u^2 \rho \, - \, 2 \frac{\phi^2}{u^4(u^{-2} - u)}\rho \ ,
\end{eqnarray}
the asymptotic expansion
\begin{equation}
\rho  \, = \, \tlmu - \tlrho \,u + \ldots \ \mathrm{and} \ \ \phi \, = \, \tilde{\zeta}_1 u + \tilde{\zeta}_2 u^2
\end{equation}
 and the near-horizon solution
 \begin{equation}
 \rho \, = \, - q (u-1) + \ldots \ \mathrm{and} \ \ \phi \, = \, \phi_0 \, + \, \order(1-u)^2 \ .
 \end{equation}
Furthermore, consistency and stability apparently requires to set either $\phi_1 = 0$ or $\phi_2=0$. This gives rise to spontaneous symmetry breaking, if one sets e.g. ${\zeta}_1 =0$ and considers ${\zeta}_1$ at the same time as a source.
\subsubsection{Flat scalar}\label{fpert}
As we discussed in section \ref{setup}, we have two scalars at hand that we can use for the symmetry breaking, the one corresponding to the separation of the probe branes in the $z$ direction along the D3 branes, and the one on the sphere separating the probe branes from the D3's, corresponding in the usual picture to a finite quark mass \cite{conpaper,fancycon,long,johanna,ingo,robdens,robchem} or chiral symmetry breaking \cite{KarchQuantum,KarchBKT,EvansBKT,EvansDiagram}. For now, let us choose the former case in which the Lagrangian in the action \reef{appactsub} becomes 
  \begin{equation}
 \sqrt{g}\mathcal{L} \, = \, \frac{1}{2} \rho'^2 \, - \, \frac{1-u^4}{2 u^4}\psi'^2 \, + \, \frac{\rho^2 }{2u^4(1-u^4)}{\psi}^2 \ .
 \end{equation}
 The equations of motion are then
 \begin{eqnarray}
0 & = & \partial_u^2 \rho \, - \,  \frac{\psi^2}{u^4(1 - u^4)}\rho \  \mathrm{and} \\
0 & = & \partial_u\left((u^{-4} - 1) \partial_u \psi\right)\, + \,  \frac{\rho^2}{u^4 - 1}   \psi \ .
\end{eqnarray}
Now there is a significant difference to the usual S-wave model, since there are two more powers of $u^{-1}$ inside the second (gauge-covariant) derivative of $\phi$. Consequently, the asymptotic scalings are different and read now
\begin{equation}\labell{zbdycond}
\rho  \, = \, \tlmu - \tlrho\,u + \ldots \ \mathrm{and} \ \ \psi \, = \, 
c_{-} \rho^{-5/2}\left( (3+u^2 \rho^2)\,\sin \, u\rho \, - \, 3 u\rho\, \cos\, u\rho\right) \, + \, \ldots
 \, =  \, \tilde{\zeta}_\psi u^5  \, + \, \ldots\ 
\end{equation}
and we define $\bar{\zeta}_\psi = \frac{\tilde{\zeta}}{\tlrho^4}$.
In principle, there is one another solution, where $\rho$ is approximately divergent, $\rho \sim e^{\psi /u}$ and $\psi$ is a decaying rapidly oscillating mode, 
\begin{equation}
 \psi \sim  c_{+} \rho^{-5/2}\left( (3-u^2 \rho^2)\,\cos \, u\rho \, + \, 3 u\rho\, \sin\, u\rho\right) \ ,
 \end{equation}but we discard this solution as it should be non-normalizable.
The near-horizon solutions are
\begin{equation}
\rho \, = \, - q (u-1) + \ldots \ \mathrm{and} \ \ \psi \, = \, \psi_0 \, + \, \order(1-u)^2 \ .
\end{equation}
\myfiguredd{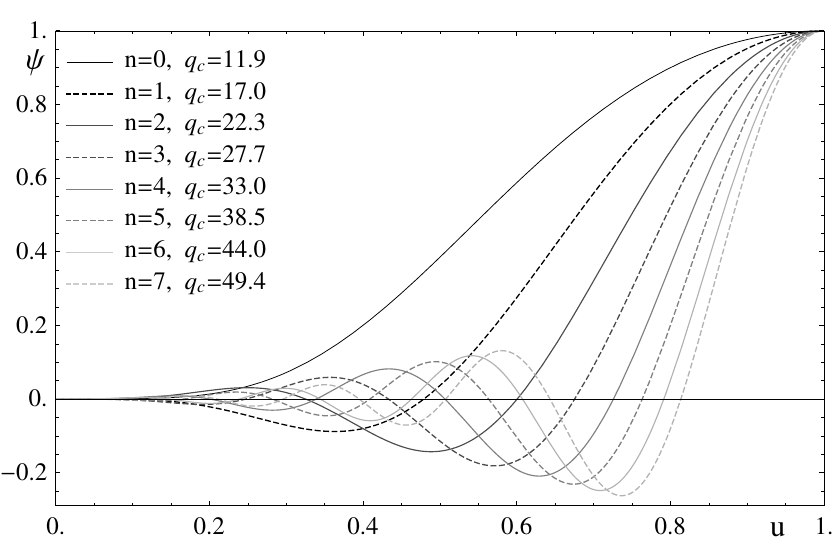}{The solutions for $\psi(u)$ close to the critical temperature, normalized to $1$.}{modes}{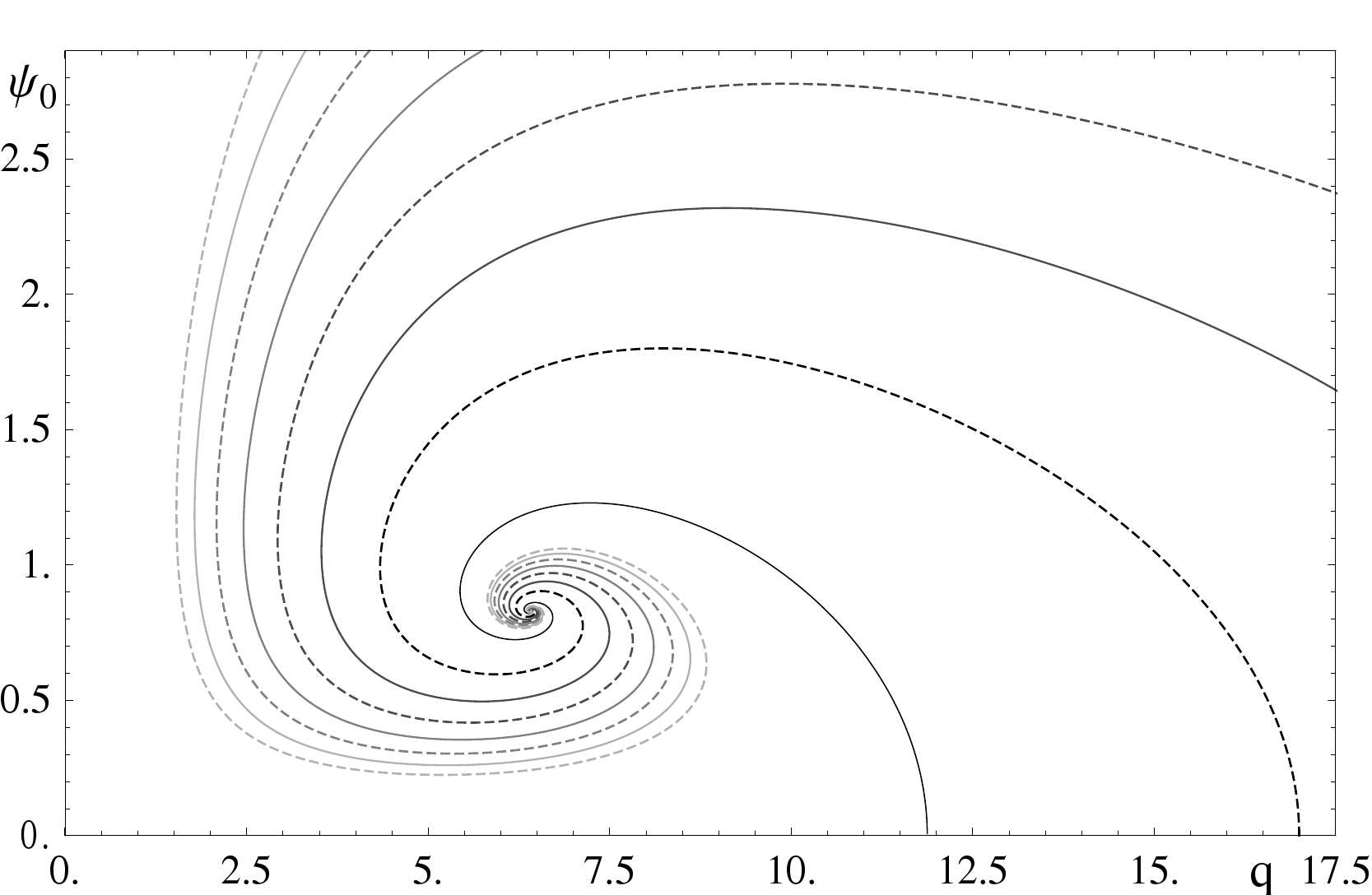}{The solution space (lines) of the internal parameters. The lines correspond to the different numbers of modes, as in the other figures. The values for $q_c$ are now $\{11.9,\ 17.0, \ 22.25, \ 27.65, \ 33.0, \ 38.5, \ 43.95, \ 49.4,\ 54.9, \ \ldots \}$}{attrac}
 
To find solutions, we first tune $\psi_0$ to a small value and study the linearized equations of motion,
\begin{equation}
 0 \, = \, \partial_u\left((u^{-4} -7 1) \partial_u \psi\right)\, + \,  \frac{q^2 (1-u)^2}{u^4 - 1}   \psi \  \mathrm{and} \ \ \rho = q (1-u)
\end{equation}
 and tune $q$ such that we satisfy the asymptotic boundary condition. We find that there are slowly oscillating solutions for $\psi$ with $n \in \mathbb{Z}^+_0$ modes at various critical $q_c$ that we show in fig. \ref{modes}.
 
In terms of the internal parameters $q$ and $\psi_0$, we find an interesting attractor behavior that we show for in fig. \ref{attrac}, as they all flow in an in-spiraling behavior to $(q,\psi_0) \sim (6.409100,0.8294795)$.
It seems that every valid solution, at least in the first quadrant of this phase space, that matches the boundary condition \reef{zbdycond} will flow to this point in the limit of small temperatures, and because of the symmetry of the action, every quadrant will have such an attractor point.
\myfiguredd{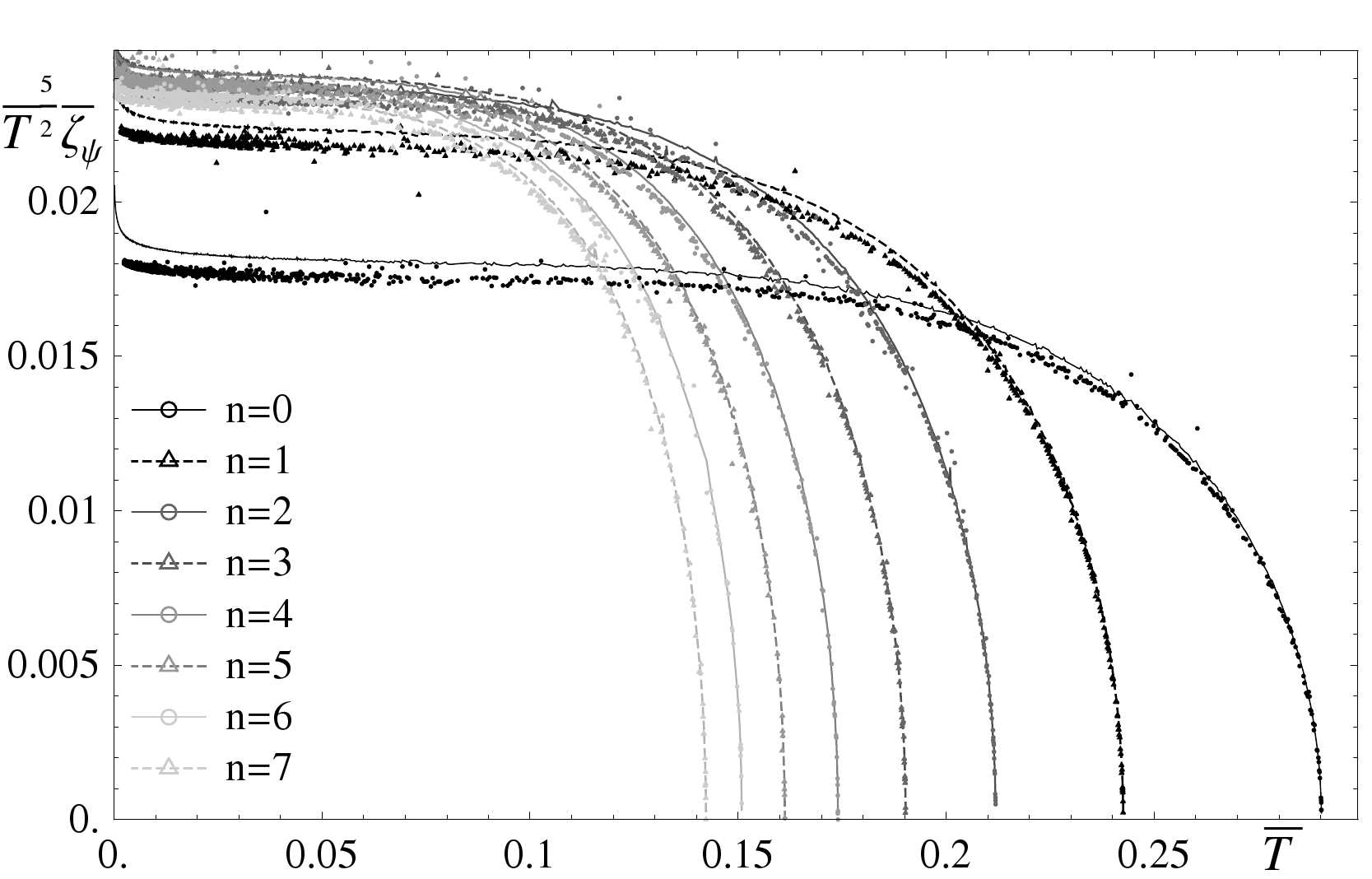}{The value of the condensate for the odd-mode solutions are negative, so the absolute value is shown here. The connected line corresponds to values obtained at $u\sim 0.1 \mu^{-4/5}$ and the dots to $u\sim 0.05 \mu^{-4/5}$.}{condval}{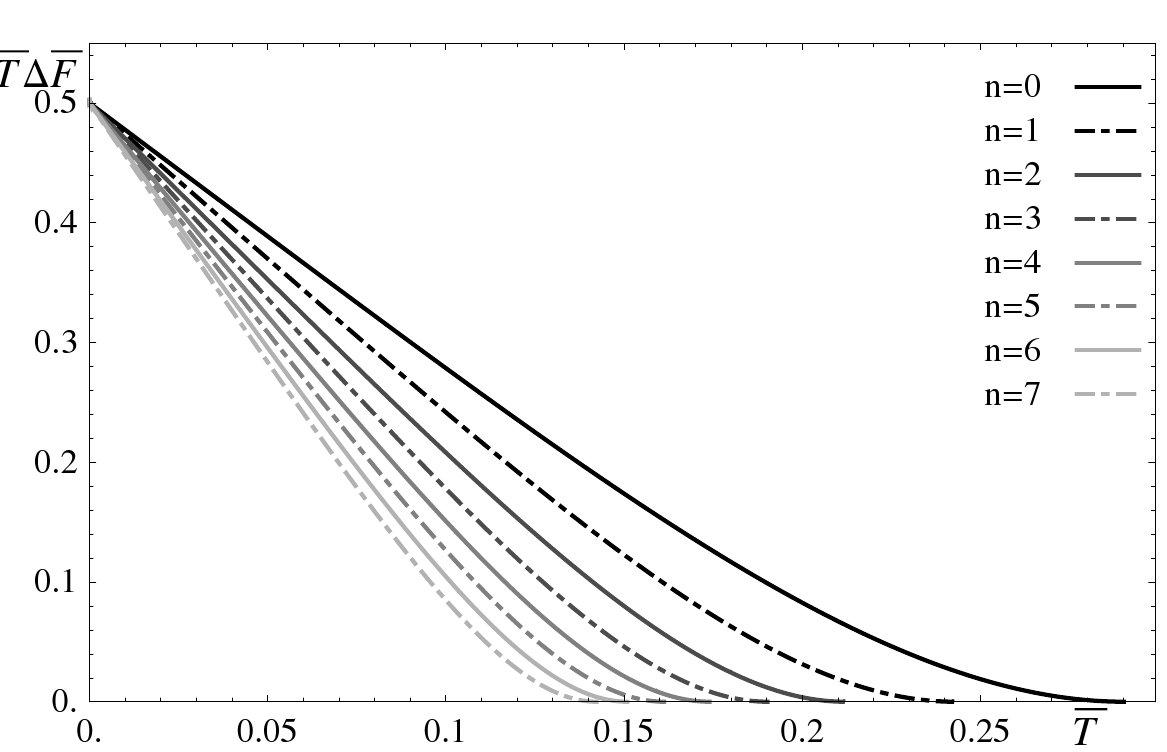}{The gap in the free energy between the unbroken and superconducting phases.}{delom}

Obtaining the value of the condensate $\zeta_\psi$ is somewhat non-trivial as this quintic mode is hard to detect. At very small values of $u$, the solution is dominated by a small numerically remnant constant term, and the range over which the quintic term is dominant is very small as the solution quickly levels off into the ``finite-$u$'' regime. The most reliable method is to fit both asymptotic modes up to $u\sim 0.1 \tlmu^{-4/5}$. Still, there is some numeric noise, 
and the results in the 0 and 1-mode solutions have already $\sim 2\%$ systematic error. To suppress the noise, we consider only values for which the coefficient of the constant term is sufficiently small. The results illustrating this are shown in fig. \ref{condval}. The result for the 1-mode solution remains below the 0-mode solution independent of systematic errors. 
An interesting observation is that we had to plot $\baT^{5/2}\bazet_\psi$ to show a finite value at small temperatures, so the condensate diverges proportional to $\baT^{-5/2}$.  This however is no surprise, since we are considering an incomplete action that can give rise to singular solutions, and this behavior should disappear when one considers the full DBI action.
  
Finally, we look at the energy gap in the free energy in fig. \ref{delom}. We find that we have indeed a second order phase transition with a continuous derivative of the free energy at the onset of the symmetry breaking solutions, and the 0-mode solution is the preferred one. There are two puzzling aspects though. Firstly, we have plotted $\bar{T} \Delta \bar{F}$, rather than just $\bar{F}$ which means that at small temperatures, the gap in the free energy is actually diverging, which is clearly not physical. Also this divergence should disappear if one considers the full DBI action. Secondly, the value of the energy gap is approaching the same value as $T\rightarrow 0$. Maybe this is related to the attractor behavior, or it may also be some restoration of supersymmetry at vanishing temperature.
%
%
\subsubsection{Compact Scalar}\label{cpert}
Now, let us consider the second scalar $\Phi^{\sin\theta}$, corresponding to the separation of the probe branes from the D3 branes.
Taking hence only the fields $\phi(u)$ and $\rho(u)$ in the action \ref{appactsub} to be non-vanishing, the relevant lagrangian is
  \begin{equation}
 \sqrt{-g}\mathcal{L} \, = \, \frac{1}{2} \rho'^2 \, - \, \frac{1-u^4}{2 u^2}\phi'^2 \, + \, \frac{\rho^2 }{2 u^2(1-u^4)}\phi^2\, + \,  u^4 \phi^2 \ .
 \end{equation}
Hence, the equations of motion are:
  \begin{eqnarray}
0 & = & \partial_u^2 \rho \, - \,  \frac{\phi^2}{u^2(1 - u^4)}\rho \  \mathrm{and} \\
0 & = & \partial_u\left((u^{-2} - u^2) \partial_u \phi\right)\, + \,  \frac{\rho^2}{u^2(u^4 - 1)}   \phi \, + \, \frac{2}{u^4} \phi \ .
 \end{eqnarray}
In their structure they are very similar to the equations \reef{hartnolleom} of the usual holographic superconductors \cite{hartnollsmall,hartnollbig} and differ just by the factors and by the power of $u$ in the blackening factor -- the latter since we are considering the induced metric on the probe brane in an $AdS_5$ black hole background rather than a planar $AdS_4$ black hole. 
 
The asymptotic scaling is now just as in \cite{hartnollsmall,hartnollbig}
\begin{equation}
\rho  \, = \, \tlmu - \tlrho\,u + \ldots \ \mathrm{and} \ \ \phi \, = \, \tilde{\zeta}_1 u + \tilde{\zeta}_2 u^2
\end{equation}
and the near-horizon expansions are 
\begin{equation}
 \rho \, = \, - q (u-1) + \ldots \ \mathrm{and} \ \ \phi \, = \, \phi_0\left(1\, + \, \frac{1}{2} (u-1)\right) \, + \, \order(1-u)^2 \ .
 \end{equation}
 Now there are two possibilities to consider either the operator corresponding to $\zeta_1$, as a source, or the one corresponding to $\zeta_2$ and require them to vanish accordingly. Again, we can find the critical values of $q$ by solving the equations of motion for small values of $\phi$,
 \begin{equation}
 0 \ = \ \partial_u\left((u^{-2} - u^2) \partial_u \phi\right)\, + \,  \frac{q^2 (1-u)^2}{u^2(u^4 - 1)}   \phi \, + \, \frac{2}{u^4} \phi \ , \ \ \rho  \, = \, q(1-u) \ ,
 \end{equation}
 for modes with $\left. \partial_u \phi(u)\right|_{u\rightarrow 0} = 0$ or $\left. \partial^2_u \phi(u)\right|_{u\rightarrow 0} = 0$, respectively.
The scalar field is then just a decaying oscillating mode as shown in fig. \ref{spheremode}.
\myfigured{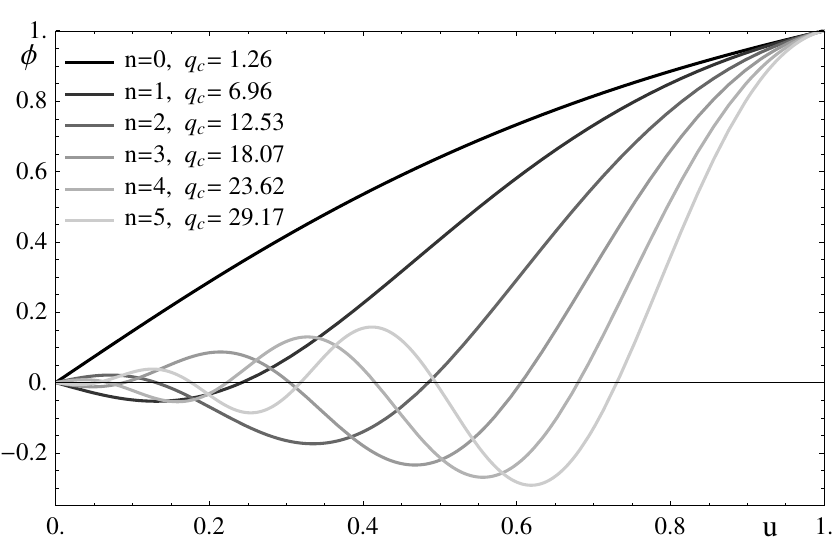}{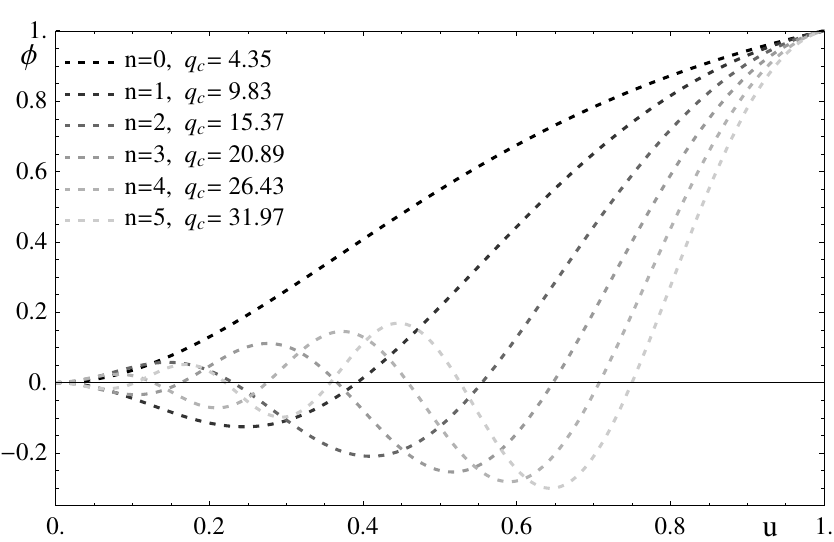}{The scalar $\phi$ as a function of the radial coordinate for the first few modes close to the critical temperature; normalized to $1$. Left: The modes with $\zeta_2 =  0$. Right: The modes with $\zeta_1 =  0$.}{spheremode}
\myfigurew{width=0.98\textwidth}{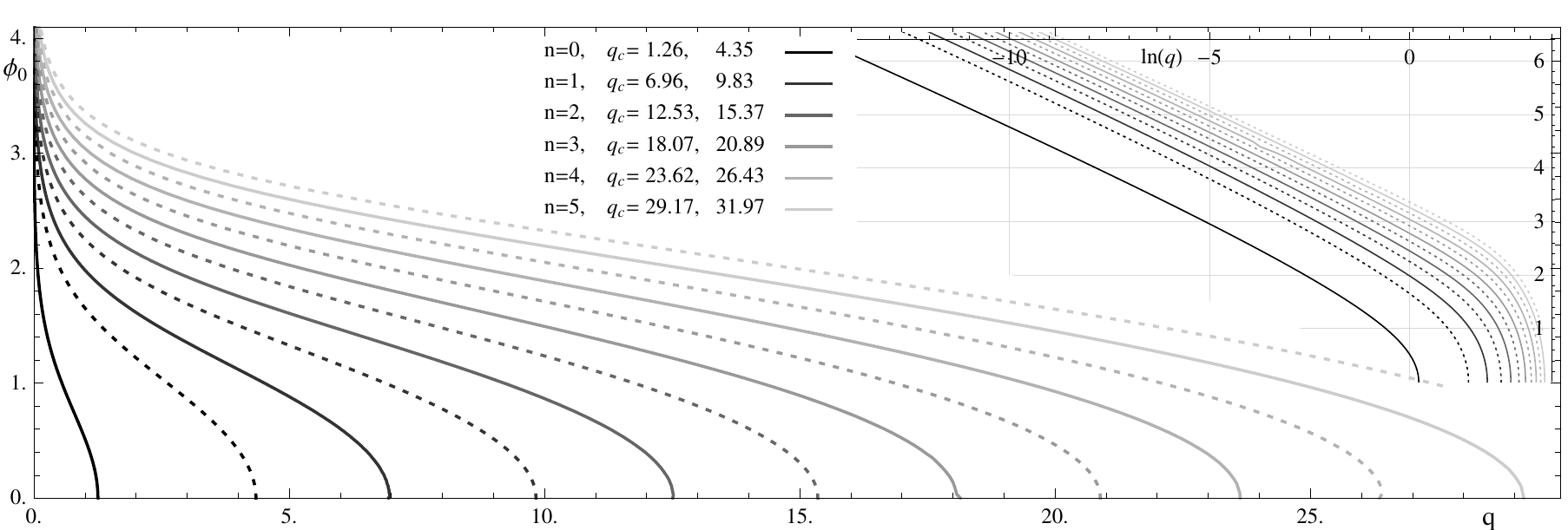}{The internal parameters $\phi_0$ and $q_c$ for the first few mode solutions. The values for $\zeta_1 =  0$ are shown dashed. The inset shows a logarithmic-linear plot.}{attrac_sphere}
\myfiguredd{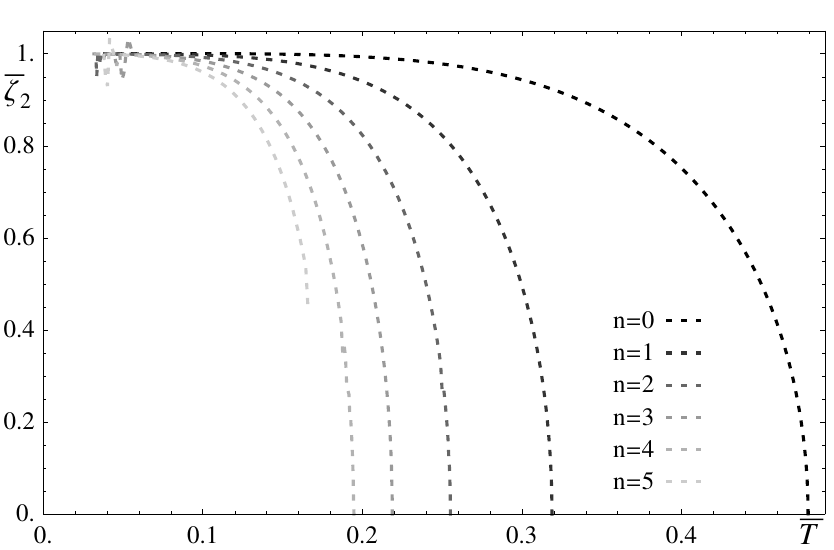}{The value of the condensate $\zeta_2$ as a function of temperature.}{condevalsphere2}{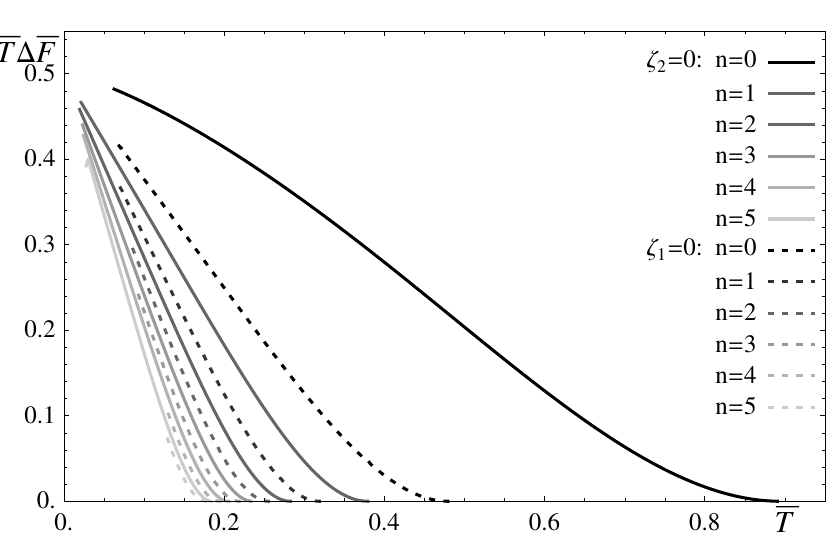}{The superconducting energy gap for the first few modes.}{engapsphere}

Then, we can increase the value of the scalar on the horizon, $\phi_0$, and track the corresponding value of $q_c$, as shown in fig. \ref{attrac_sphere}. Now, find again a kind of attractor behavior with a divergence $\phi_0  \sim - 0.33 \ln q + const.$, where the constant increases with increasing modes.
%
\myfigured{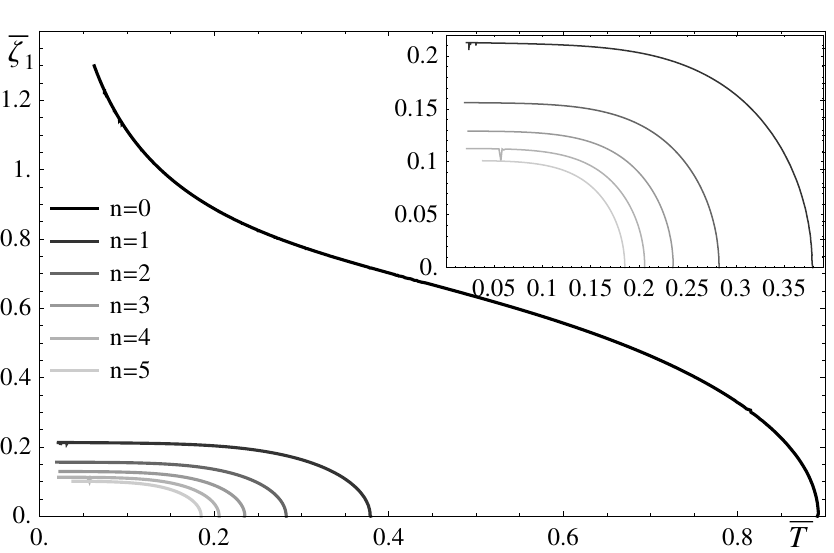}{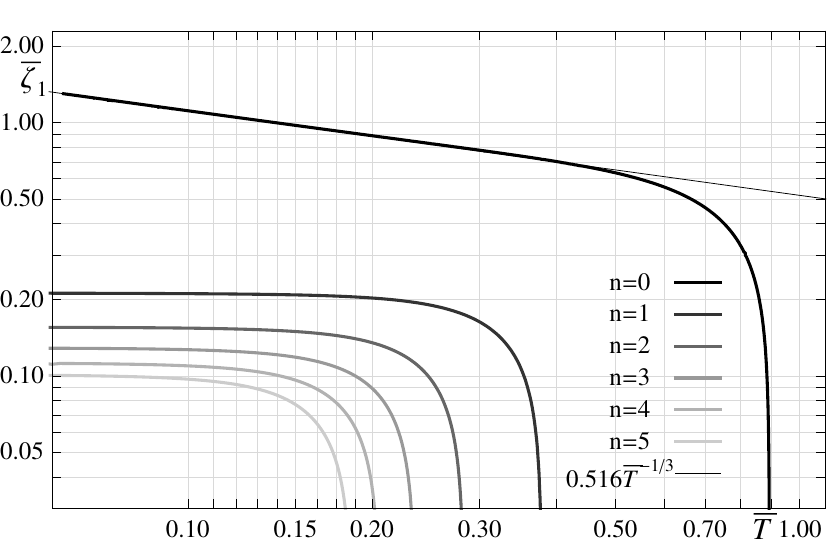}{The value of the condensate $\zeta_1$ as a function of temperature as a linear and double logarithmic plot. Again, the value of the condensate is negative for odd $n$ and only the absolute value is shown.}{condevalsphere}

Looking at the result in terms of the physical parameters in figs. \ref{condevalsphere}  and \ref{condevalsphere2}, the condensate $\bazet_1 = \frac{\tilde{\zeta}_1}{\sqrt{\tlrho}}$ or $\bazet_2 = \frac{\tilde{\zeta}_2}{\tlrho}$ and the temperature $T$, we find a picture that is qualitatively similar to the previous case -- but with much larger critical temperatures. Now, only the $0^{th}$ mode of the $\zeta_2=0$ case diverges, proportional to $T^{-1/3}$. Interestingly, $\zeta_2$ flows precisely to $\pm 1$ for all modes at small temperatures.

Finally, we can again look at the energy gap in fig. \ref{engapsphere}, where we see clearly again a second order phase transition, and the fact that the $\zeta_2=0$, $n=0$ mode is the thermodynamically preferred phase. We find again the divergence of the energy gap at small temperatures, and the flow to $\baF = 1/(2 \baT)$ at small temperatures. 
%
%
%
\subsection{Comparing the Condensates}\label{apert}
\myfigure{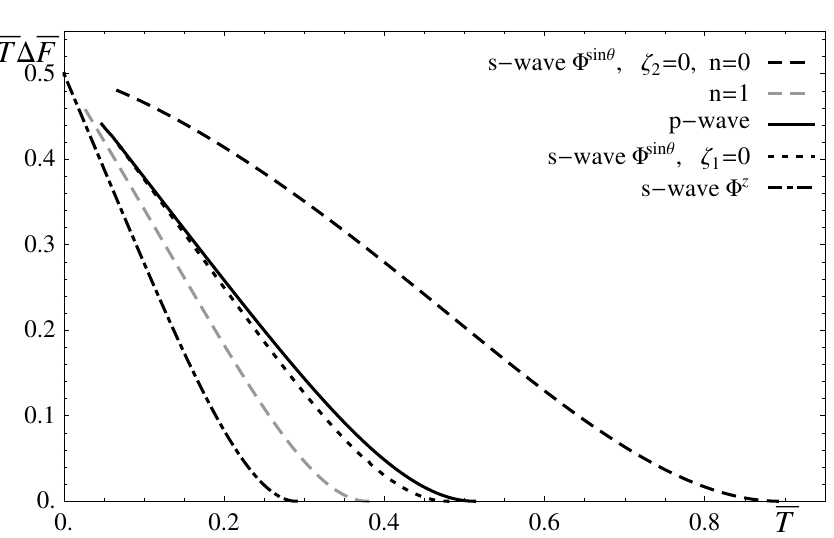}{Energy gap comparing the different condensates.}{compall}
Now, let us compare the different superconducting phases to determine which one is the stable one. Using again the canonical ensemble, we see in fig. \ref{compall}, that the dominant condensate is always the s-wave $\zeta_1$.
For comparison, we also show the $n=0$ mode of the p-wave, and it turns out that this is only metastable and should decay to the s-wave. 

Interestingly, the p-wave and the $\zeta_2$ s-wave condensates behave very similarly.

Certainly, we assumed here that the canonical ensemble is relevant as fixing the particle number is a sensible thing to do for superconductors. For completeness, we also provide some results related to the grand canonical ensemble that may be relevant in QCD contexts in app. \ref{gcapppert}. 
 
%
\section{Full DBI Action}\label{fullact}
Now, let us consider the ``full'' DBI Action. Certainly the non-abelian DBI Action \reef{robbigact} is not complete and receives corrections at higher order in $\alpha'$ \cite{Kitazawa,Koerber,Bachas,Wijnholt}. In any event, so far it is not possible to compute even the non-abelian DBI action exactly because of the symmetrized trace. The symmetrized trace prescription requires us to take a symmetrized average over all orderings of $\cd_\mu \Phi$, $F_{\mu \nu}$ (and generically $[\Phi^a,\Phi^b]$).
Hence, in oder to obtain an exact solution, we would need to do some resummation, which will probably be possible only in the simplest cases. In order to still preserve the features of the regularity of the DBI action and the geometry of the $S^5$ sphere, we will do the following: 

Defining the symmetrization operator $\s$, we can write the DBI action schematically as
\begin{equation}
S \ = \ - \int \str \sqrt{\cdot} \ = \ - \int \tr \s \sqrt{\cdot} \ = \ - \int \tr \s \sqrt{\s (\cdot)}  ,
\end{equation}
where in the second identity we used $\s(\s(f)\s(g)) = \s(fg)$ and assumed that we can expand the square root.
Even inside the square root, one can go as far as symmetrizing any pair of the ``atomic'' expressions.
The ``approximation'' that we will do is to evaluate this ``first'' symmetrization in the matrix representation \reef{paulidef}, i.e. we will replace any pair of such objects with their anti-commutator. Since the action \reef{braneaction} contains these terms only at even order, the expression inside the square root will even be proportional to the identity, and one can readily compute the result.

That this procedure is not exact can be seen at the following example: $\s(\tau_1 \tau_2 \tau_1 \tau_2) = \s(\tau_1 \tau_1 \tau_2 \tau_2)$ because of the symmetrization, and we still have $\s(\s(\tau_1 \tau_2) \s(\tau_1 \tau_2)) = \s(\s(\tau_1 \tau_1) \s(\tau_2 \tau_2))$. If we fallaciously evaluate the anticommutators in each side using the representation \reef{paulidef}, the left hand side will vanish, but the right hand side will be proportional to the identity.
As our ``approximation'' is however identical to the so-called symmetrized trace prescription that has been used in the literature \cite{johannascon7big} and it preserves the features of regularity and the spherical geometry, we will use it for the rest of the paper.
Because of the commutator term $[\Phi^a,\Phi^b]$ in \reef{robbigact}, let us not write down the action now, but postpone this to the separate cases.
The induced metric simplifies in this ``approximation'' to
\begin{eqnarray}\nonumber
ds^2 &  = &  2 \frac{L^2}{ u^2} \tau_0 \left(-\left((1-u^4)+ \rho^2 \psi^2 + u^2\frac{\rho^2 \phi^2}{1-\phi^2} \right)  d\tlt^2 + d\vec{\tx}_2^2 \right. \\ 
& & ~~~~~~~~~~~~~~~~\left. + \left(\frac{1}{1-u^4} + \psi'^2 + \frac{u^2 \phi'^2}{1-\phi^2}\right) du^2  \right)\, + \, 2 \tau_0 L^2 (1-\phi^2) d\Omega_2^2  \ .
\end{eqnarray}
 
Before we start finding the solutions in the broken phases, let us find the solution of the unbroken phase.
In this case, the action in the parametrization of \reef{elansatz} becomes after an integration over the sphere 
\begin{equation}
S \ = \ - \frac{1}{2\pi^3}\sqrt{\lambda}N_c\int d^4 \tilde{\sigma} \frac{1}{u^4}\sqrt{4 - u^4  \rho'^2} \  .
\end{equation}
Note that this is a somewhat unusual normalization of the gauge field, arising from the definition in \reef{elansatz}. 
The solution for $\rho$ is
\begin{equation}\label{rhoblanksol}
\rho'  \, = \, \frac{2 q}{\sqrt{1+q^2 u^4}} \ , \ \ \rho \, = \, 2\sqrt{iq}\left(F(i\sinh^{-1}\sqrt{iq},i) - F(i\sinh^{-1}\sqrt{iq}u,i) \right) \ ,
\end{equation}
where the parameter $q$ is related to the isospin density through $\tlrho = 2 q$ as there are now no sources. $F(\cdot,\cdot)$ is the elliptic integral of the first kind. At large values of $\tlrho$, this gives us $\tlmu \sim \frac{\sqrt{\tlrho}}{2 \sqrt{2 \pi}}\Gamma\left(\frac{1}{4}\right)^2 q^{3/2}$, whereas at small values, we have $\tlmu \sim \tlrho$, as expected.
The value of the action becomes 
\begin{equation}
S \ = \ - \frac{1}{2\pi^3}\sqrt{\lambda}N_c \tilde{V}_{2+1} \times \frac{2}{3} \left(q\sqrt{iq}F(i\sinh^{-1}\sqrt{iq},i) -\sqrt{1-q^2} \right)\ ,
\end{equation}
where we included the usual counter term $\frac{1}{3}\sqrt{\gamma}$ and introduced a volume factor $\tilde{V}_{2+1}$ for the dimensionless ``flat'' coordinates, that gives rise to a factor of $2\pi^3 T^3$ in the densitized euclidean action. In the limit of large $q$, this expression evaluates to $S \sim - \frac{\sqrt{\lambda}N_c \tilde{V}_{2+1}}{2\pi^3} \frac{1}{6 \sqrt{\pi}}\Gamma\left(\frac{1}{4}\right)^{2}$ and at small $q$, it becomes just the usual conformal $S = -\frac{\sqrt{\lambda}N_c \tilde{V}_{2+1}}{2\pi^3} \frac{2}{3}$.
 \subsection{P-Wave}\label{pfullact}
Now, let us consider only the ``magnetic'' field $\omega(u) \tau_1 d\tilde{x}$ as in section \ref{ppert}.
The action becomes now in the parametrization of \reef{vectoransatz} and after an integration over the sphere
\begin{equation}
S \ = \ \frac{\sqrt{\lambda}N_c }{2\pi^3}\int d^4 \tilde{\sigma} \frac{1}{u^4}\sqrt{4 + u^4 \left( (1-u^4)\omega^2 - \rho'^2\right) - u^4\frac{ \rho^2 \omega^2}{1-u^4}} \ ,
\end{equation}
such that equations of motion for $\rho(u)$ and $\omega(u)$ become
\begin{eqnarray}
\partial_u \frac{\rho'}{\sqrt{4 + u^4 \left( (1-u^4)\omega^2 - \rho'^2\right) - u^4\frac{ \rho^2 \omega^2}{1-u^4}}} &= &\frac{ \rho \omega^2}{(1\!-\!u^4)\sqrt{4 + u^4 \left( (1\!-\!u^4)\omega^2 - \rho'^2\right) - u^4\frac{ \rho^2 \omega^2}{1-u^4}}}\, ,~~~~~~~ \\
\!\!\! \partial_u \! \left( \! \frac{\omega' (1-u^4)}{\sqrt{4 + u^4 \left( (1\! -\! u^4)\omega^2 - \rho'^2\right) - u^4\! \frac{ \rho^2 \omega^2}{1-u^4}}}\right)\!\! &= &\! - \frac{ \rho \omega^2}{(1\!-\!u^4)\sqrt{4 + u^4 \left( (1\!-\!u^4)\omega^2 - \rho'^2\right) - u^4\!\frac{ \rho^2 \omega^2}{1-u^4}}} \, .~~~~~~~
\end{eqnarray}
Near the horizon, the boundary conditions can be found by an expansion of the equations of motion around $u=1$ as
\begin{equation}
\rho  \ = \  \frac{2 q}{\sqrt{1+q^2 }} (1 - u) \, +\, \ldots \  \mathrm{and} \ \   
 \omega \ = \  \omega_0 \, - \, \frac{\omega_0}{64} \left(\frac{2 q}{\sqrt{1+q^2 }} \right)(1 - u)^2
\, + \,   \ldots  \ ,
\end{equation}
where as usual the vanishing of the ``electric potential'' on the horizon is required  by regularity. We chose the parametrization of the electric field analogous to the solution \reef{rhoblanksol} for the gauge field in the unbroken phase in order to ensure regular solutions and also for practical purposes for the fine-tuning near $\frac{ q}{\sqrt{1+q^2 }} = 1$. Also physically, $q$ still represents the flux through the horizon, or the isospin density that is not carried by the condensate.
The asymptotic expansions are the same as in the perturbative case \reef{ominf}, $\rho  = \tlmu - \tlrho \,u + \ldots$ and $\omega  = \tilde{\zeta}_\omega\, u +\ldots$.

\myfigured{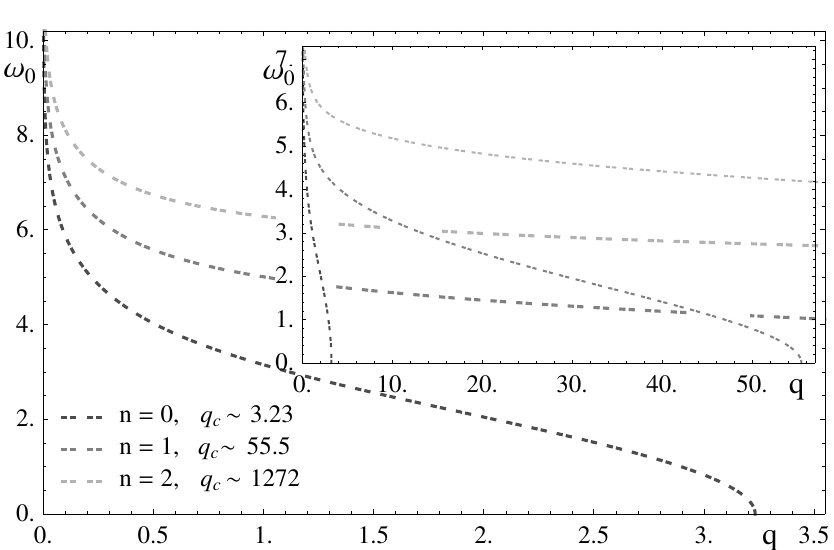}{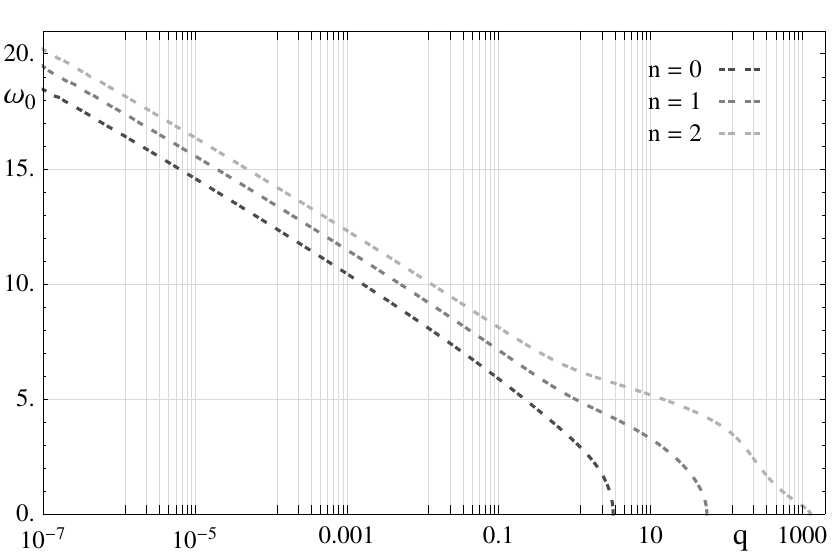}{The relation of the internal parameters for the first few modes of the spontaneously symmetry-broken solutions for the p-wave condensate.}{omfullattrac}
In fig. \ref{omfullattrac}, we look at the relation of the internal parameters $\omega_0$ and $q$ for the first three modes of the spontaneous symmetry breaking solutions. As usual,
the value of $q$ decreases with an increasing horizon value of the symmetry breaking field. Now however, because of the ``shape'' of the DBI action, the critical values of $q$ are separated by a large factor rather than just equally-spaced. The critical temperatures are correspondingly suppressed, at values $\baT_c \sim \{0.39327, 0.094872, 0.019068, \ldots \}$. The divergence of $\omega_0$ at small temperatures is logarithmic with $\omega_0 \sim - 0.87\ln q + const.$ or expressed in another way, the value of $q$ is exponentially suppressed at large $\omega_0$. This just reflects the fact that most of the density is carried by the condensate in that limit.
\myfigured{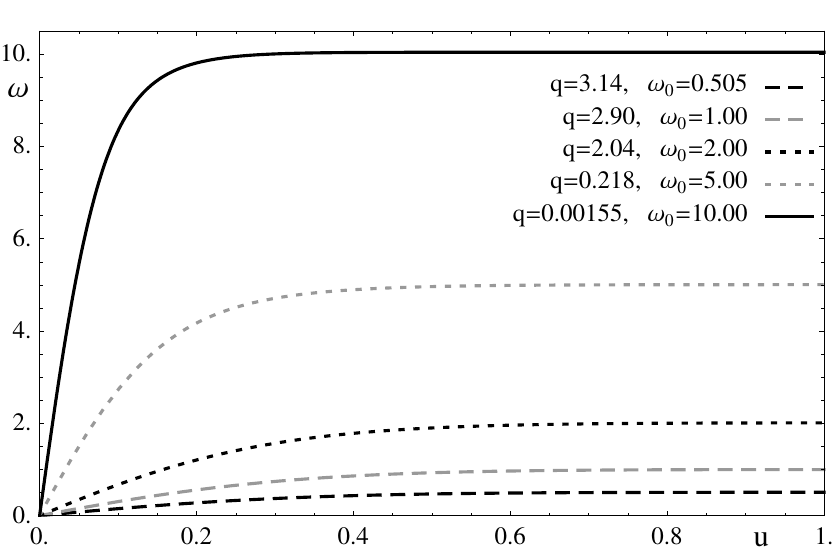}{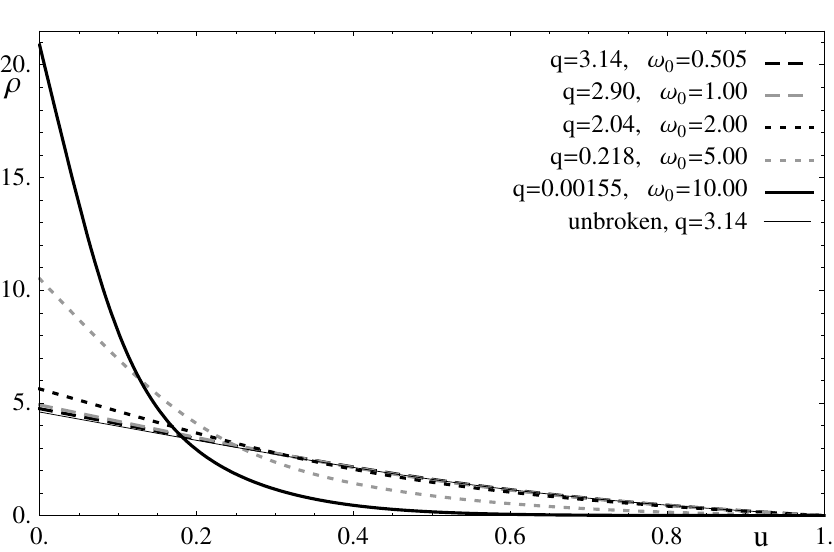}{The $n=0$ mode solution of the p-wave superconductor for different choices for the boundary condition: Symmetry breaking field (left) and isospin potential (right). The black narrow line shows the isospin potential for $q=3.14$ for reference. Curves for larger values of $q$ will be {\it below} this curve.}{omfullcondprofiles}

The profiles of $\omega(u)$ and $\rho(u)$ are shown in fig. \ref{omfullcondprofiles} for some selected values of the 0-mode solution. We find the usual ``positive feedback'' between the symmetry breaking field and the isospin potential. 
This also illustrates how the isospin chemical potential or density is supported by the symmetry breaking field:
While the profile for the isospin potential for large values of $q$ and small values for $\omega_0$ is similar to the one in the unbroken phase (as it should), for small values of $q$ and large values of $\omega_0$ the potential (and its ``electric field'') are almost entirely sourced by the charged field, which in turn is approximately constant up to large radii (or small values of $u$).
\myfiguredd{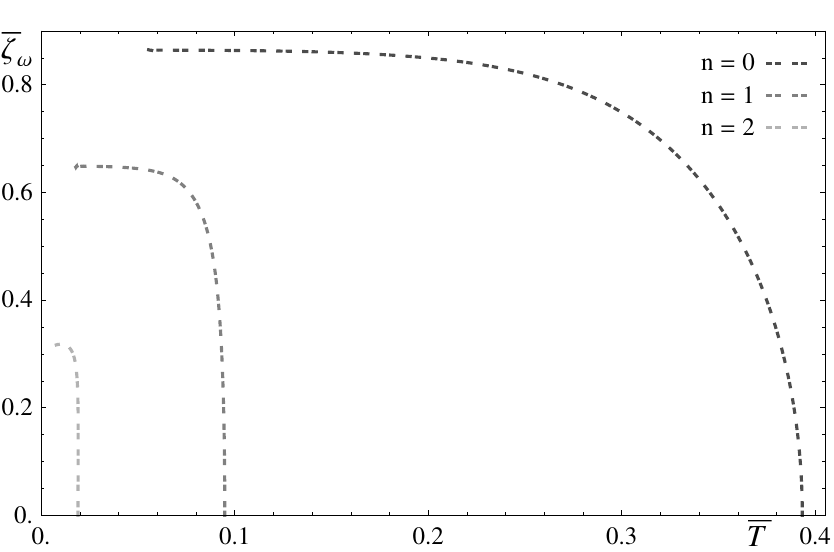}{Absolute value of the superconducting condensate for the first few modes of the p-wave superconductor.}{omfullcond}{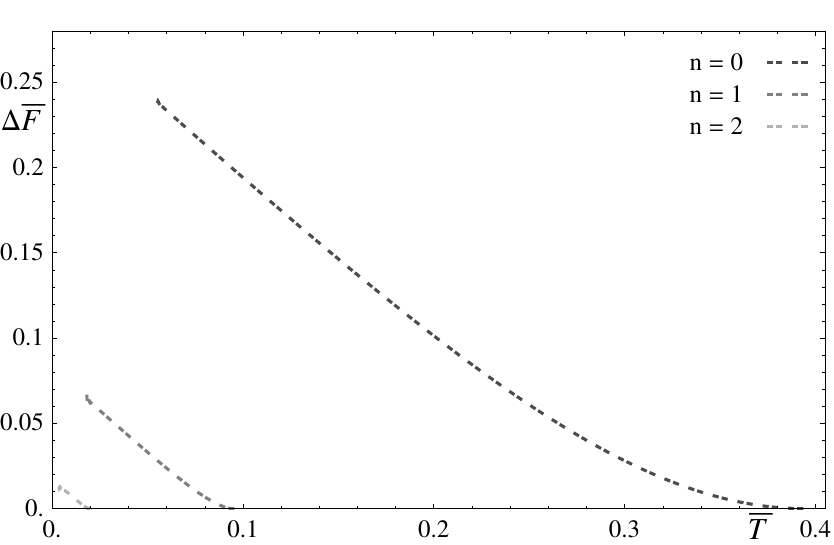}{Gap in the free energy with respect to the unbroken phase.}{omfullgap}

In fig. \ref{omfullcond}, we show the (absolute) value of the symmetry breaking condensate $\bar{\zeta}_\omega \, = \, \frac{\tilde{\zeta}}{\sqrt{\tlrho}}$ in the broken phase. While the critical temperatures for the $n^{th}$ mode scale approximately proportional to $4^{-n}$, the small-temperature limit of the condensate rather saturates -- both are due to the form of the DBI action.

In contrast to this behavior, the curves of the energy gap in fig. \ref{omfullgap} are approximately invariant under an approximate scaling with the critical temperature. Also in contrast to the $\alpha'^2$ expansion, the energy gap is finite at small temperatures. This is a clear feature of the regularity of the DBI action.
\subsection{Compact Scalar}\label{cfullact}
Next, let us consider the s-wave condensate from the ``compact'' scalar that describes the separation of the D5 probes on the $S^5$. In contrast to the case of the $\alpha'$ expansion, this will contain some more interesting physics, as we have now black hole and non-blackhole embeddings in which the probe branes do not extend down to the horizon, just as in the usual massive probe brane embeddings (see e.g. \cite{long,johannaphase,robbig,fancytherm}). This may give us an additional first oder phase transition.

In the parametrization of \reef{scalaransatz}, the action becomes now
\begin{equation}
 S \ = \ - \frac{\sqrt{\lambda}N_c }{2\pi^3} \int d^4 \tilde{\sigma}\frac{1}{u^4}(1-\phi^2)\sqrt{4 \! \left(\! 1- u^2\frac{\rho^2\phi^2}{(1\!-\! \phi^2)(1-u^4)}\right)\left(1+u^2(1\!-\!u^4)\frac{\phi'^2}{1\!-\!\phi^2}\right) - \rho'^2 u^4} \ ,
\end{equation}
which gives us the equations of motion
\begin{eqnarray}\nonumber\label{compeq}
\!\!\!\!\!\!\!\!& & \!\!\!\!\!\!\! \partial_u \frac{\rho'(1-\phi^2)}{\sqrt{4  \left(1- u^2\frac{\rho^2\phi^2}{(1-\phi^2)(1-u^4)}\right)\left(1+u^2(1-u^4)\frac{\phi'^2}{1-\phi^2}\right) - \rho'^2 u^4}} \\ 
\!\!\!\!\!\!\!\!& & ~~ =\ \frac{4 \rho\phi^2\left((1-\phi^2)+u^2(1-u^4)\phi'^2\right)}{u^2(1-u^4)(1-\phi^2)\sqrt{4  \left(1- u^2\frac{\rho^2\phi^2}{(1-\phi^2)(1-u^4)}\right)\left(1+u^2(1-u^4)\frac{\phi'^2}{1-\phi^2}\right) - \rho'^2 u^4}} ~~~ \\ \nonumber
\!\!\!\!\!\!\!\!& & \!\!\!\!\!\!\!  \partial_u \frac{4  \left((1-\phi^2)(1-u^4)- u^2\rho^2\phi^2\right)\phi'}{u^2 (1-\phi^2)\sqrt{4  \left(1- u^2\frac{\rho^2\phi^2}{(1-\phi^2)(1-u^4)}\right)\left(1+u^2(1-u^4)\frac{\phi'^2}{1-\phi^2}\right) - \rho'^2 u^4}}  \\ 
\!\!\!\!\!\!\!\!& & ~~~ = \ - \phi\frac{ 4\left(u^2 \rho^2\frac{1-2\phi^2}{1-u^2} + 2 (1-\phi^2) \right)  +  4 u^2 \phi'^2\left(1-u^4 + u^2\rho^2 \right) - u^4 \rho'^2(1-\phi^2) }{u^4(1-\phi^2)\sqrt{4  \left(1- u^2\frac{\rho^2\phi^2}{(1-\phi^2)(1-u^4)}\right)\left(1+u^2(1-u^4)\frac{\phi'^2}{1-\phi^2}\right) - \rho'^2 u^4}} \ .
\end{eqnarray}
The asymptotic embeddings are again $\rho   =  \tlmu - \tlrho\,u + \ldots $ and $ \phi  =  \tilde{\zeta}_1 u + \tilde{\zeta}_2 u^2 + \ldots$, where either $\zeta_1$ or $\zeta_2$ is considered as a source and hence required to vanish, and the other one is considered to be the condensate. Again, we define $\bazet_1 = \frac{\tilde{\zeta}_1}{\sqrt{\tlrho}}$ and $\bazet_2 = \frac{\tilde{\zeta}_2}{\tlrho}$. For the ``IR'' boundary condition, we have to consider the black hole embeddings and the non-blackhole embeddings separately.
\subsubsection{Black Hole Embeddings}\label{bhfullact}
The usual near-horizon expansion around $u = 1$ gives us 
\begin{equation}
\rho  \ = \  \frac{2 q}{\sqrt{1+q^2 }} (1 - u) \, +\, \ldots \  \mathrm{and} \ \   
 \phi \ = \  \phi_0 \left(1 \, + \,  \frac{ (u-1)}{2({1+q^2 })} \right)
\, + \,   \ldots  \ ,
\end{equation}
where we chose again the parametrization in terms of $q$ to obtain a regular solution and reflect the flux sourced by the black hole.
\myfigured{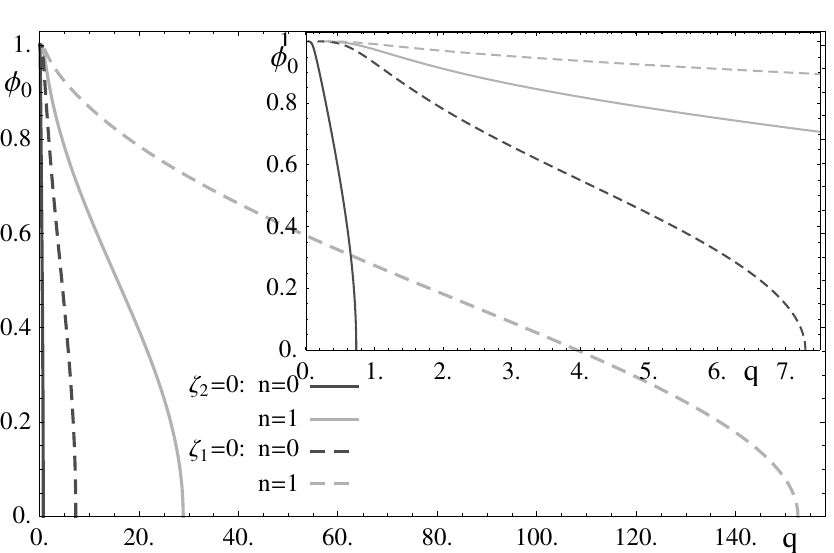}{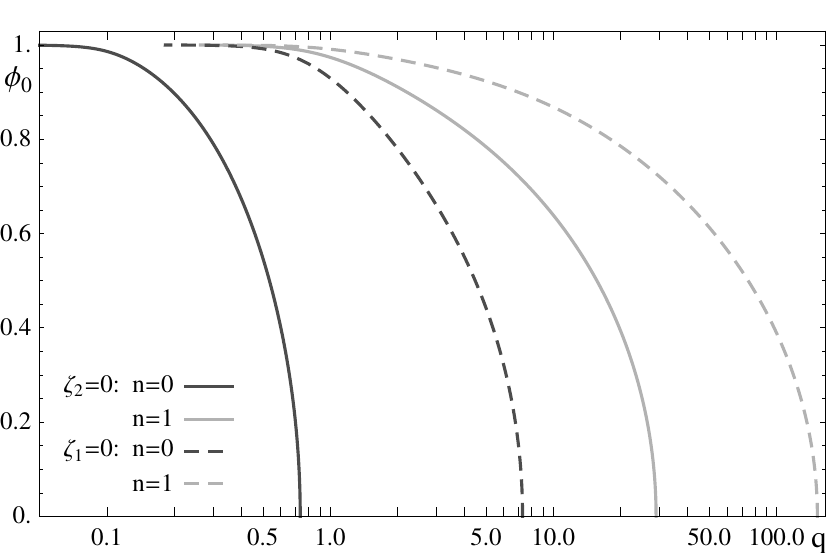}{The horizon value of the scalar field as a function of the ``isospin flux' through the horizon as a linear and a linear-logarithmic plot for the $n=0$ and $n=1$ modes.}{psitbhfullattrac}

Looking at the ``IR'' parameters for the $n=0$ and $n=1$ mode solutions in fig. \ref{psitbhfullattrac}, we find again that the ordering of the critical values for $q$ is the same, and they are exponentially separated, rather than linearly as at order $\alpha'^2$. There is an attractor at $q=0$ and $\phi_0 = 0$, which just reflects the geometry of the sphere. It tells us that all of the charge density becomes supported by the condensate as the embedding approaches the critical embedding in which the brane touches the horizon at only one point. Intuitively, this arises because an embedding with finite flux out of the horizon would give a singular value for the action at this point and is hence avoided by the DBI action. 
\subsubsection{Minkowski Embeddings}\label{minkfullact}
Now, let us look at the case where the branes do not extend to the horizon, but the black hole is located ``in between'' the branes.
In order to find the boundary condition at the point $u_0$ at which the size of the $S^2$ sphere vanishes, $\phi \rightarrow 1$, we substitute the Ansatz 
\begin{equation}
\phi \, = \, 1 \, - \, \phi'_0 (u_0-u^\alpha) \ \ \mathrm{and} \ \ \rho \, = \, \mathfrak{q} (u_0-u)^\beta
\end{equation}
into the equations of motion \reef{compeq}. In any case we should assume $a, b>0$ in order to obtain a sensible solution with the given Ansatz, and furthermore we will assume the limitation of all the possible permutations of 
\begin{equation}\label{expcond}
\{2\beta-2, \alpha-1,0 \} \, < \, \{2-\alpha,4\beta-2-\alpha,2\alpha-1,2\beta-1\} \ . 
\end{equation}
Note that the solution for the abelian case, $b=1/2$ and $a=1$ violates that bound marginally. Under the condition \reef{expcond} we get now the four leading orders
\begin{eqnarray}
0 & = & u_0^4 \mathfrak{q}^2\beta(u_0-u)^{2\beta -2} \left(2\alpha^2 - \alpha\beta -2\beta^2 \right) \, - \, 4 \alpha^2 \phi'_0 (u_0-u)^{2\beta -2} \left(3 -u_0^4 \right) \, + \,  \left(8 + 4\alpha \right) \ \ \mathrm{and} \\
0 & = &  (u_0-u)^{\beta +\alpha -2} \left(2\alpha^2 - \alpha\beta -2\beta^2 \right) \ .
\end{eqnarray}
The non-trivial solutions to these equations are $\beta=1$, $\phi'_0 = \frac{3}{3 u_0-u_0^5}$ and $\alpha = -\frac{1}{4}(1 \pm \sqrt{17})$, such that the only solution that also satisfies $\alpha >0$ (and the bound \reef{expcond}) is
\begin{equation}\label{embscale}
\phi \, \sim \, 1 \, -\frac{3}{3 u_0-u_0^5}(u_0-u) \ \ \mathrm{and} \ \ \rho \, \sim \, \mathfrak{q} (u_0-u)^{ \frac{\sqrt{17} -1}{4}} \, . 
\end{equation}
It was argued in the abelian case of the D3-D7 system in \cite{robdens} (and for the D3-D5 system in \cite{fancytherm}) that the D3-Dp strings that are necessary to source the flux in the Minkowski embedding of the probe branes would be in disbalance with the tension of the Dp-branes at $u_0$ and cause the Minkowski solution to collapse into a black hole solution. However, it turns out that the electric flux 
\begin{equation}
\frac{\delta \mathcal{L}}{\delta \partial_u A_t} \, \propto \, \frac{(1-\phi^2)^2 \rho' \sqrt{1-u^4}}{\sqrt{4\left((1-u^4)(1-\phi^2)-\rho^2\phi^2\right)\left(1-\phi^2+\phi'^2(1-u^4)\right)-\rho'^2(1-u^4)}} 
\end{equation}
does vanish in our case near $u \sim u_0$ as
\begin{equation}\label{fluxscale}
\frac{\delta \mathcal{L}}{\delta \partial_u A_t} \, \propto \,  \mathfrak{q}\frac{\sqrt{17} - 1}{4 u_0 \sqrt{1-u_0^4}}\sqrt{\frac{6}{3 u_0-u_0^5}}\times (u_0-u)^{\frac{\sqrt{17} +1}{4}} \ .
\end{equation}
This just means that the flux on the probe brane is entirely sourced by the scalar condensate. Hence, the parameter $\mathfrak{q}$ is now related to the density of Dp-Dp strings near $u_0$ rather than the number D3-Dp strings in the abelian case.
In fact, one can consider the variation of the flux, $\partial_u \frac{\delta \mathcal{L}}{\delta \partial_u A_t}$ to describe the density of the source-Dp-Dp strings, and from \reef{fluxscale} we see that it scales as $\mathfrak{q}(u_0-u)^{\frac{\sqrt{17} -3}{4}}$. As $\frac{\sqrt{17} -3}{4}\sim 0.28$ and hence the density vanishes as $u\rightarrow u_0$, this is consistent with the fact that the scaling of the probe brane embedding, \reef{embscale}, does not depend on the value of $\mathfrak{q}$. 
\myfigured{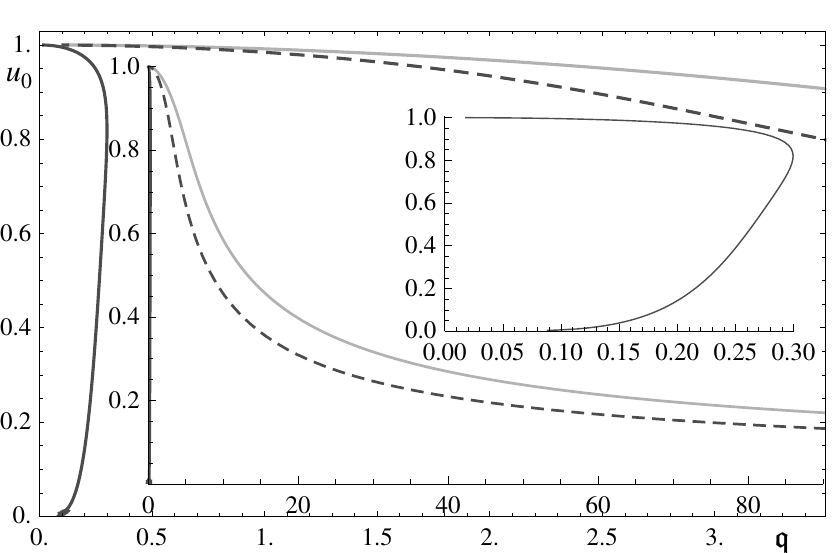}{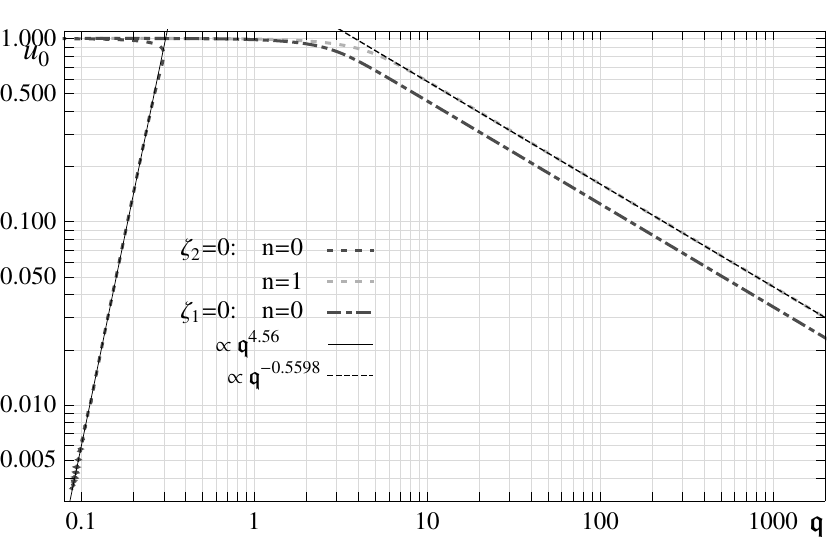}{The horizon value of the scalar field as a function of the Dp-Dp string density parameter $\mathfrak{q}$ as a linear and a double-logarithmic plot for the $n=0$ and $n=1$ modes in the Minkowski embedding.}{psitminkfullattrac}

To discuss the dependence of $\mathfrak{q}$ on $u_0$, we have to keep in mind that for large $u_0$ close to $1$, the brane profile has some non-trivial ``trumpet-like'' shape (i.e. it extends almost down to the black hole), and for small values of $u_0$, it spans almost ``straight'' from $u_0$ to $0$ (i.e. it is approximately flat, at large radii). This tells us precisely why the value of $\mathfrak{q}$ that we need to obtain the 0-mode solution for $\phi_2 = 0$ vanishes as $u_0\rightarrow 0$ as shown in fig. \ref{psitminkfullattrac}: The attraction from the black hole is small, such that the boundary condition is almost matched by the geometric part of the DBI action and one needs little flux to enforce the boundary condition. Some more consequences of this will appear later.
However for the higher mode solutions and for $\phi_1 = 0$, $\mathfrak{q}$ becomes large as we need a large flux to support the profile, also because the ``space'' that is available in the inverse radial coordinate $u$ to ``bend' the brane vanishes. It is somewhat surprising though that the scaling exponents around $u_0 \rightarrow 0$ seem to be non-rational. Because of our definition of $\mathfrak{q}$ that does not contain the other factors in \reef{fluxscale}, $\mathfrak{q}$ vanishes $\propto \sqrt{1-u_0^4}$ as $u_0\rightarrow 1$.

While in the black hole embedding phase the isospin density is also carried an appropriate number of free quarks and anti-quarks that are in equilibrium with ``mesons'', now all the isospin density is carried by mesons.
\subsubsection{Combined Results}\label{combfullact}
\myfigured{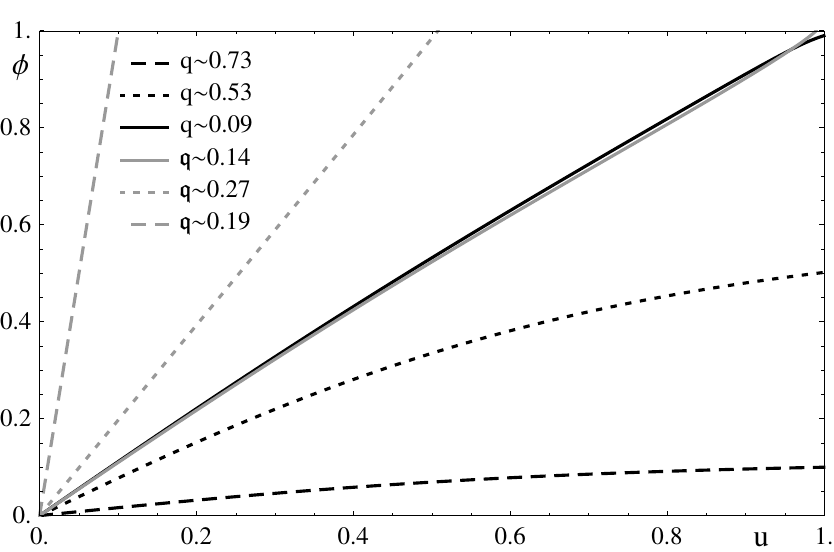}{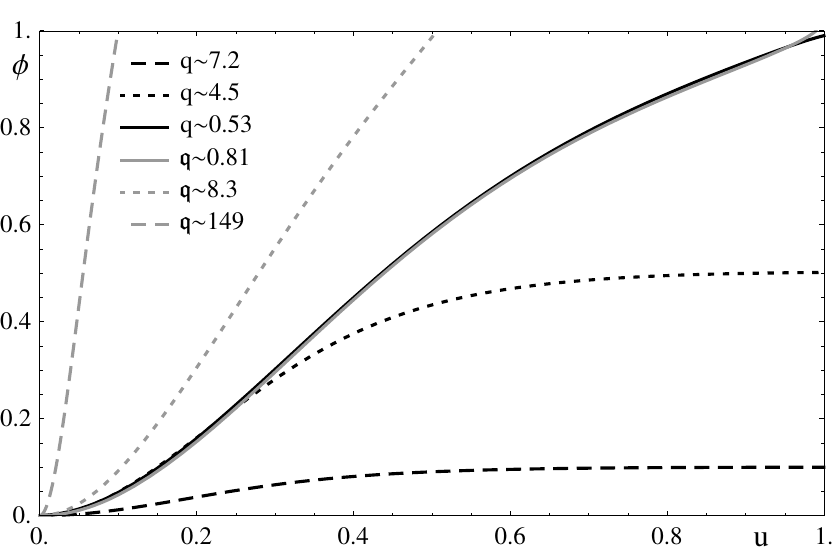}{The 0-mode profiles of the probe brane embeddings for both Minkowski and black hole solutions for $\zeta_2=0$ (left) and $\zeta_1=0$ (right).}{psitfullprofiles}
In fig. \ref{psitfullprofiles}, we show the brane profiles of the 0-mode solutions for the two different condensates for a few values of $\phi_0$ and $u_0$, respectively. We see that the profiles for $\phi_0$ and $u_0$ close to $1$ actually cross close to the horizon, and also the curves for $\phi_0 \sim 0.5$  and  $\phi_0 \sim 1$ are similar in the UV (small $u$). These features are reminiscent of the meson melting phase transition \cite{long,fancytherm} and its many aspects/relatives in different regimes (see e.g. \cite{KarchBKT,EvansBKT}).
\myfigured{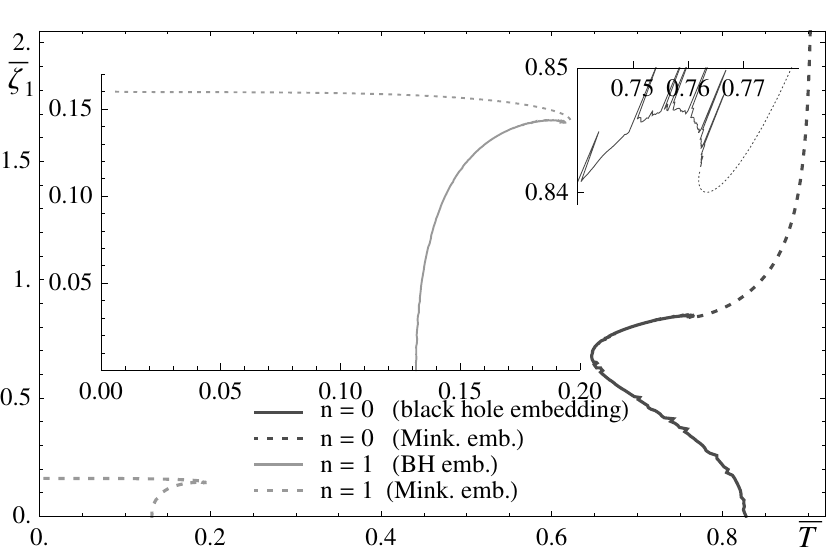}{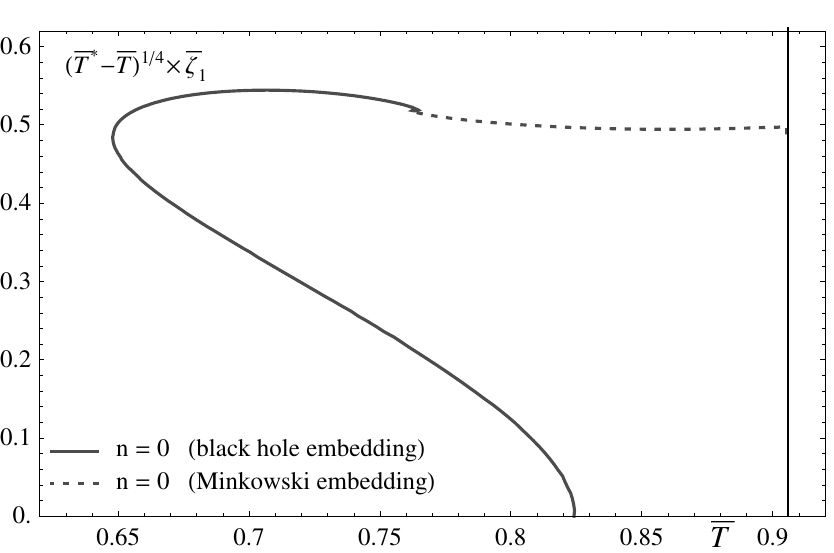}{The value of the condensate $\zeta_1$ for the first two modes. The insets on the left show details of the $n=1$ mode and the crossover between the black hole and Minkowski embedding. The plot on the right shows the condensate scaled by $(T^* - T)^{1/4}$ for some $\baT^* = 0.905956$ that is indicated by the black line.}{psit1fullcond}

Just as in the case of the $\alpha'^2$ expansion in section \ref{pert}, the $n=0$ mode is again different from the other modes, as we show in fig. \ref{psit1fullcond}. Now, it starts off again proportional to $\sqrt{T_c-T}$, and there follows a multivalued behavior that is typical for black hole - Minkowski transitions. However rather than extending to $\baT=0$, the phase ``ends'' at some finite temperature $\baT^*$ at which the value of the condensate diverges, approximately proportional to $(\baT^* - \baT)^{-1/4}$ as we show on the right plot. 
From the gravity side, this is not surprising, as at vanishing $u_0$, $\zeta_1$ naturally diverges as $\tilde{\zeta}_1 \sim 1/u_0$. Since the scale that sets the temperature is just given by the isospin density, i.e. by the ``electric flux'', and we only need some finite flux to support this embedding as discussed above, the phase ends at this finite $\baT^*$. Usually the phases extend to $\baT=0$, where the divergence of the condensate is canceled by its temperature scaling due the to dimensional factor. We will discuss the relevance of this effect further below.

The $n=1$ mode starts rather unusually proportional to $\sqrt{T-T_c}$ -- probably a saturation effect due to the regularity of the DBI action. Then it has the transition to the mesonic phase, which extends to $T=0$. The fact that the value of the mesonic phase is almost constant is qualitatively not surprising, as in the mesonic phase most the free quarks have already condensated. While it is certainly still not impossible that there is a smooth density of D3-Dp strings that can potentially condensate, these stretch very far and are hence energetically disfavored
%
%
\myfiguredd{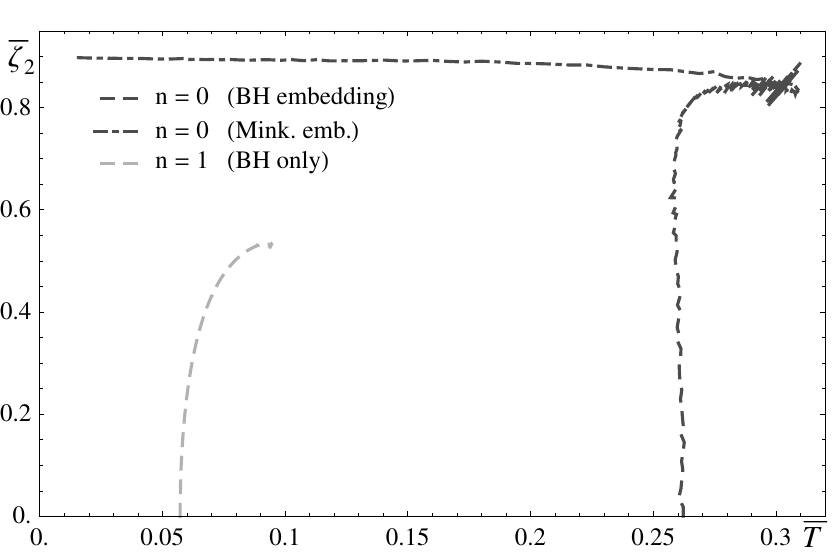}{The value of the condensate $\zeta_2$ as a function of temperature. For the $n=1$ phase, we could only find the black hole embedding.}{psit2fullcond}{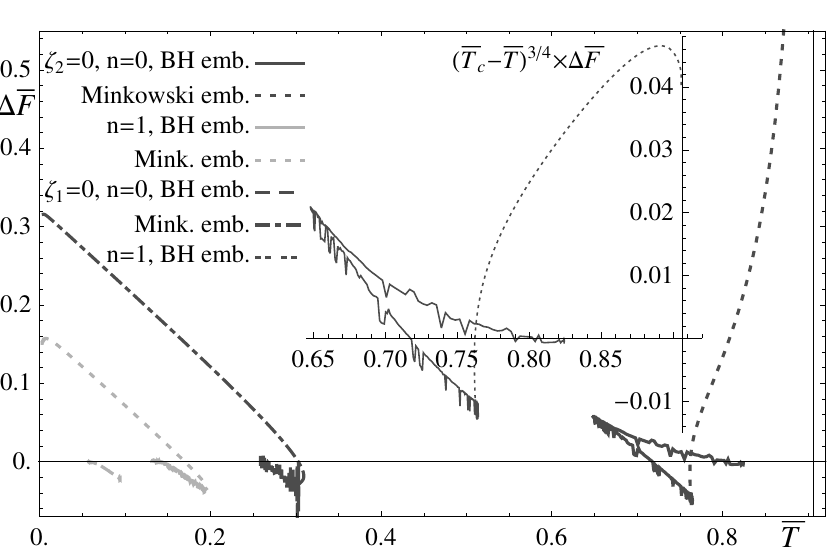}{The gap in the free energy for the first few modes of the ``compact'' s-wave solution. The inset shows the energy gap scaled by $(\baT^* - \baT)^{3/4}$.}{psitfullen}

For $\varphi_2 = 0$, we could find for the $n=1$ mode only the black hole embedding, as the Minkowski embedding was beyond reach of the numerics for most values of $u_0$. The results are shown in fig. \ref{psit2fullcond}, and we find that now already the $n=0$ mode is similar to the $n=1$ mode of the other scalar. This is because one needs a large charge density to ``bend'' the brane profile, and this is also sourced in regimes away from the UV. Overall, we also see that the results get somewhat noisy, especially close to the critical embedding at $u_0=1$ or $\phi_0=1$. 

Looking at the energy gap for the Helmholtz free energy in the canonical ensemble in fig. \ref{psitfullen}, we see again that the $\varphi_2 = 0$, $n=0$ phase does not extend out to $T=0$, but it diverges at $\baT^*$. The power is not quite clear and can range from $-0.66$ to $-0.75$, but if one wants to have the same $\baT^*$ as in the diverging condensate, values close to $-0.75$ seem to give better fits. The ``drop'' that we see in the right plot in fig. \ref{psitfullen} can be due to a subleading term that is proportional to $(\baT^* - \baT)^{1/4}$. The ratio of the divergences of the energy gap and the condensate is then $3$, consistent with the ratio of their dimensional scaling. The geometrical reason is the large volume factor of the probe branes that extend only to some small $u_0$ (i.e. large minimum AdS radius), so it is perfectly consistent, and the problem again only arises because the flux needed to support this embedding is finite, rather than diverging. This divergence seems to be an artifact of the adapted symmetrized trace prescription, as we will discuss in the conclusions in section \ref{conclusions}. Before the divergence, we see a second order phase transition and some multivaluedness, which is in a similar form also seen in the usual black hole - Minkowski transitions. Rigorously speaking, the first turning point would give rise to a $0^{th}$ oder phase transition, which would not appear if the Minkowski phase would extend to vanishing temperatures as this part of the black hole would not be realized and we would have a first order phase transition to the Minkowski phase before.

In the higher modes, the superconducting phase transitions ``reverse'', such that the black hole embeddings are always metastable and we rather have a direct first order phase transition to the Minkowski phase. The change of the order of the phase transition is qualitatively not surprising, as in the black hole phase there is an equilibrium of mesons and free quarks, so the condensate can condensate smoothly, but in the mesonic phase, most of the free quarks have already condensated, so one cannot have a continuous phase transition.
 \subsection{Flat Scalar}\label{ffullact}
Finally, we consider the case of the scalar $\Phi^z$ that originates from the embedding of the probe branes in the flat (field theory) directions. Using the parametrization \reef{scalaransatz}, the action is
\begin{equation}
S \ = \ - \frac{\sqrt{\lambda}N_c }{2\pi^3} \int d^4 \tilde{\sigma}\frac{1}{u^4}\sqrt{\left(4 - \frac{\rho^2 \psi^2}{1-u^4}\right)\left(1+\psi'^2\frac{1-u^4}{4}\right)-u^4 \rho'^2} \ ,
\end{equation}
which gives us the equations of motion
\begin{eqnarray}
\partial_u \frac{\rho'}{\sqrt{\left(4 - \frac{\rho^2 \psi^2}{1-u^4}\right)\left(1+\psi'^2\frac{1-u^4}{4}\right)-u^4 \rho'^2}} & = & \frac{\rho \psi^2\left(4+\psi'^2(1-u^4)\right)}{4(1-u^4)\sqrt{\left(4 - \frac{\rho^2 \psi^2}{1-u^4}\right)\left(1+\psi'^2\frac{1-u^4}{4}\right)-u^4 \rho'^2}} \\
\partial_u \frac{\psi'\left(4(1-u^4) - \rho^2 \psi^2\right)}{4\sqrt{\left(4\! - \!\frac{\rho^2 \psi^2}{1\!-\!u^4}\right)\left(\!1+\!\psi'^2\frac{1\!-\!u^4}{4}\right)\!-\!u^4 \rho'^2}} & = & \frac{-\rho \psi^2\left(4+\psi'^2(1-u^4)\right)}{4(1-u^4)\sqrt{\left(4 - \frac{\rho^2 \psi^2}{1-u^4}\right)\left(1+\psi'^2\frac{1-u^4}{4}\right)-u^4 \rho'^2}} \ .
\end{eqnarray}
Again, we expand around $u=1$ to obtain the near-horizon limit
 \begin{equation}
\rho  \ = \  \frac{2 q}{\sqrt{1+q^2 }} (1 - u) \, +\, \ldots \  \mathrm{and} \ \   
 \psi \ = \  \psi_0 \, - \, \frac{\psi_0}{64} \left(\frac{2 q}{\sqrt{1+q^2 }} \right)(1 - u)^2
\, + \,   \ldots  \ 
\end{equation}
and the asymptotic expansions are identical to the ones in the perturbative expansion,
$\rho   =  \tlmu - \tlrho\,u + \ldots$ and $\psi = \tilde{\zeta}_\psi u^5  +  \ldots$, that we discussed in detail in section \ref{fpert}. Also, we define again $\bazet_\psi = \frac{\tilde{\zeta}_\psi}{\tlrho^2}$. As in section \ref{fpert}, there are some technical problems due to this highly suppressed scaling of $\psi$ in the UV, so in particular the results for the $n=1$ mode will be very noisy and only available in some regimes at all. This noise is why we will show the individual data points in some of the plots.
\myfigured{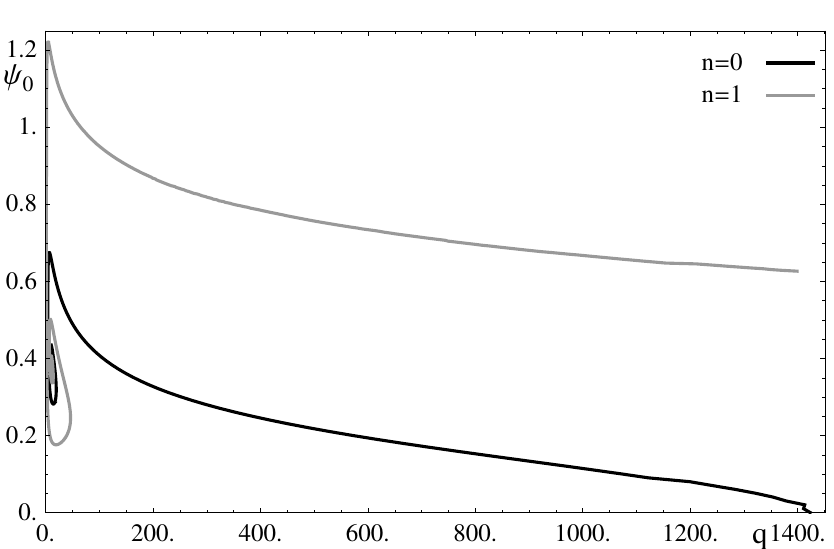}{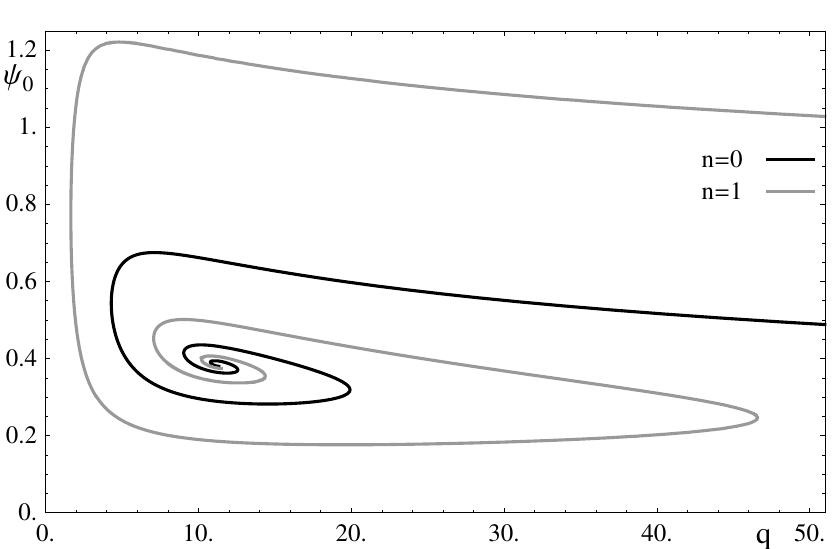}{The boundary condition for the scalar in terms of the flux through the horizon. The right plot shows the details of the attractor region at small $q$.}{zfullattrac}

First, let us look at the ``internal'' parameters, the flux out of the horizon $q$ and the ``separation'' of the branes on the horizon, $\psi_0$ in fig. \ref{zfullattrac}. We find again an attractor behavior with a fixed point around $(q,\phi_0)\sim (11.2,0.385)$, but now the ``spiral'' is heavily distorted and the critical value of $q$ for the $n=0$ mode moved to a very large value of $q_c\sim 1423$. For the $n=1$ mode the numerics could not reach far enough to obtain $q_c$.
\myfigured{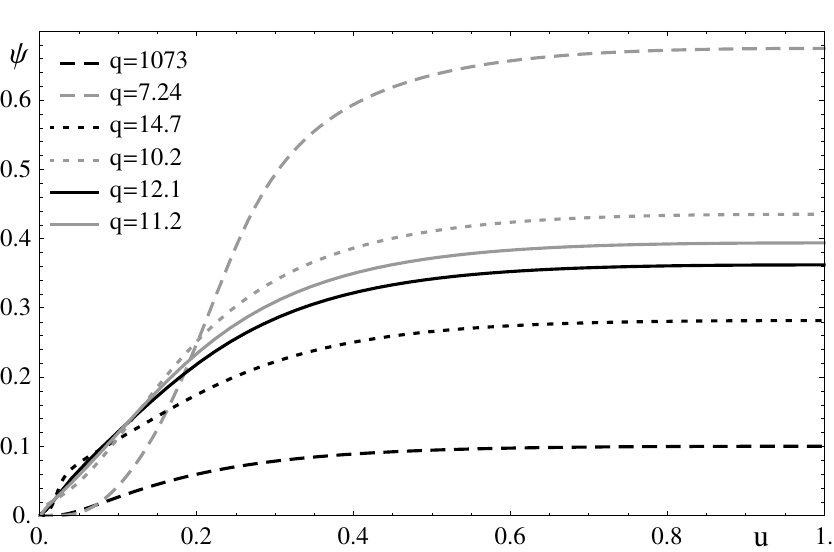}{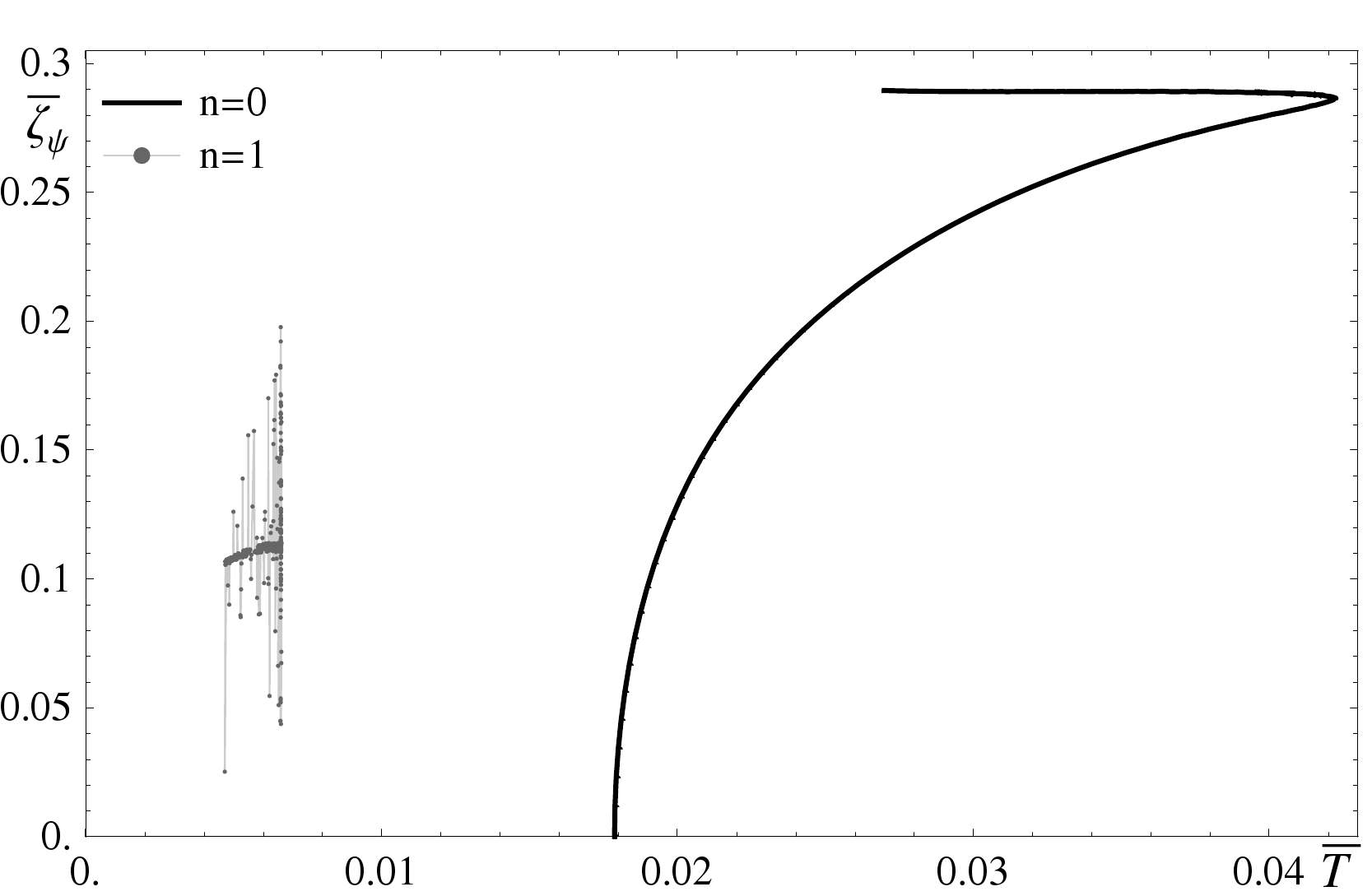}{Left: Embedding profiles for the ``flat'' scalar for the $n=0$ mode for several values of $\psi_0$ and $q$ near the maxima and minima of the attractor spiral. Right: Scalar condensate $\zeta_\psi$ as a function of temperature. For the $n=1$ mode only a small regime is shown due to numerical issues.}{zfullcondprofiles}

In fig. \ref{zfullcondprofiles}, we show on the left the profiles of $\psi$ of the 0-mode solution for choices of $q$ and $\psi_0$ near the minima and maxima of the attractor spiral. We see that the curves seem to approach some non-singular limiting profile and the large value of the condensate is obtained by following that profile more closely and shifting the $\propto u^5$ behavior further into the UV. The temperature dependence of the condensate is surprisingly similar to the higher modes of the other scalar in figures \ref{psit1fullcond} and \ref{psit2fullcond}, even though they are geometrically very different and also the internal parameters behave differently. This may be a universal feature of scalars in the DBI action. Physically this may arise because we are looking at very small temperatures, i.e. large dimensionless densities $\tlrho$, while the dimensionless flux parameter at the horizon, $q$, that describes the un-condensated isospin density quickly moves towards its finite attractor value. Hence as soon as the internal parameters are in the attractor region, most of the density has already condensated and chance the condensate is approximately constant. For the $n=1$ mode, we show only a small region in this plot, as the numerically obtained value of $\varphi_5$ becomes extremely noisy. 
\myfigure{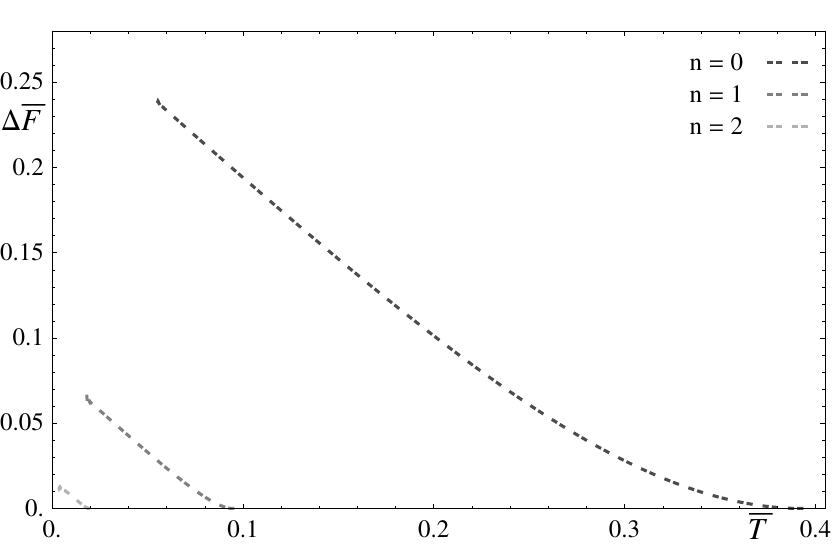}{Energy gap for the scalar condensate corresponding to the position of the probe branes in the field theory directions.}{zfullen}

As with the value of the condensate, also the energy gap, shown in fig. \ref{zfullen}, is very similar to the other scalar, albeit with a much lower critical temperature and correspondingly lower value. Again, we see that the second order phase transition ``inverts'' and we have now a first order phase transition directly to the regime in which the density has almost completely condensated. The curve for the $n=1$ mode is again very similar to the $n=0$ curve, upon a scaling with the critical temperature.
%
%
\subsection{Comparing the Phases}\label{afullact}
\myfigurerw{0.65\textwidth}{0.3\textwidth}{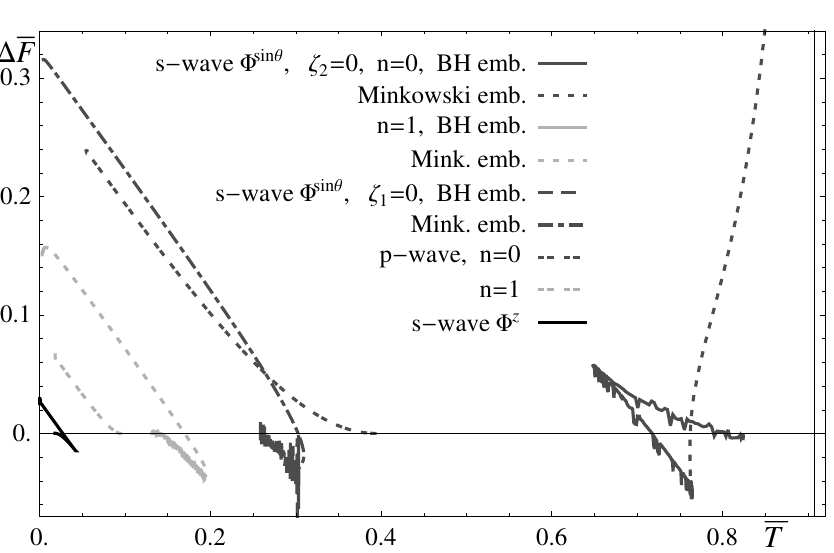}{The energy gap of the different condensates using the ``full'' DBI action.}{compfullen}
In fig. \ref{compfullen}, we compare the energy gap of the different condensates. We see first of all that the ordering of the critical temperatures does not change compared to the $\alpha'^2$ expansion, but especially the smaller critical temperatures are further suppressed. In addition to the changes discussed already in the previous subsections, we see now that there seems to be a first order phase transition between the p-wave and $\zeta_2$ s-wave phases. Obviously we studied the different phases only separately, and if one considers the full system there would be some mixing between the phases. Also it is not quite sure whether such detail would persist if one considers the rigorous symmetrized trace and the higher order corrections of \cite{Kitazawa,Koerber,Bachas,Wijnholt}. This would probably also cause the $\zeta_1$ s-wave phase to extend to $T=0$.

Further, we also see that while the $T=0$ value of the energy gap is now different for the different phases, the first derivative, and hence the amount by which the entropy is lowered by the symmetry breaking becomes 
\begin{equation}
\Delta \bar{S} \ = \ - \frac{\partial \Delta \baF}{\partial \baT} \ \sim \ 1
\end{equation}
at small temperatures. This is not contradicting the third law of thermodynamics or the positivity of entropy, as it is well know that these holographic systems describe at small temperatures a quantum liquid with non-vanishing ground state degeneracy \cite{zerosound,funnel,luttinger,paolo}. In fact, we can write down the free energy in the unbroken phase,
\begin{equation}
\baF_{unbroken} \ = \ - \frac{2}{3}\baT^3  \left( \sqrt{1-(2\baT^2)^{-2}} + \frac{1}{\baT^2}\sqrt{i/(2 \baT^2)}F(i\sinh^{-1}\sqrt{i/(2 \baT^2)},i)  \right)\ ,
\end{equation}
which gives us $\bar{S} \rightarrow 1$ as $\baT \rightarrow 0$. Hence the symmetry breaking precisely cancels the zero-temperature entropy, such that the ground state degeneracy of the broken phase vanishes. 
Inspecting the curves further for the heat capacity $\bar{c}_v = \partial_\baT (\baF+\baT \bar{S}) = \baT \partial_\baT^2 \baF$, we see that in the cases of first order phase transitions the heat capacity is increased whereas for the second oder phase transitions it is lowered.

As in the perturbative case, we show some results related to the grand canonical ensemble that may be relevant in some other context in appendix \ref{gcappfull}.
\section{Discussion and Conclusions}\label{conclusions}
In this paper, we studied the phase diagram of a system of two probe ``defect'' D5 branes in the presence of a non-trivial $SU(2)$ world-volume gauge potential in an $AdS_5\times S^5$ background that is created by a stack of $N_c \gg 1$ D3 branes. The field theory dual is a $3+1$ dimensional $SU(N_c)$, $\mathcal{N}=4$ SYM theory coupled to two flavors of charged fermions that are confined to a $2+1$ dimensional ``defect''. Certainly, the system is in any case still invariant under an overall $SU(2)_f$ rotation, but one way to interpret the system is by choosing $\rho(u) \tau_3 d\,t$ (with an inverse radial coordinate $u$) as a potential which corresponds to a finite isospin density. Having thus broken the $SU(2)$ explicitly to a $U(1)$, we considered three different ``channels'' of spontaneous symmetry breaking of this $U(1)$ using fields that do not commute with that potential (e.g. proportional to $\tau_1$). The candidates were a magnetic component of the world-volume gauge field, $\omega(u)\tau_1 d\, x$, giving rise to a p-wave superconductor, and two different scalars giving rise to s-wave superconductors: A scalar corresponding to the embedding of the probe branes in the flat (3+1) field theory directions orthogonal to the (2+1) defect directions but along the D3s, and one corresponding to the embedding in the ``internal'' $S^5$, away from the D3 branes. In some sense they could be imagined as a relative separation of the probe branes (if we had chosen the scalar proportional to $\tau_3$), but it should be rather considered as some entanglement at two relatively separated positions.

Technically, we used the non-abelian DBI action that governs the world-volume fields on the probe branes in the probe approximation $N_f \ll N_c$ and studied the symmetry breaking fields separately. As it is physically most sensible for our considerations of superconductors, we chose to work in the the canonical ensemble, where we fixed the density. For completeness, we referred to what would happen in the grand canonical ensemble, that would be relevant e.g. in a semiconductor in app. \ref{gcapp}.

Qualitatively, our setup could be motivated from condensed matter physics by considering a 2+1 dimensional system with (at least) two separate species of particles that interact in the effective field theory -- e.g. a multi-layered system like cuprate superconductors or multi-layered graphene. Depending on which kind of potential we choose, the density that we turn on corresponds then to some non-trivial combination of the particle number of either species, not only the ``isospin'' density of particles in one species and antiparticles in the other. 
In the string theory side, the symmetry breaking condensates correspond to some combination of D5-D5 strings, which translate on the field theory side into bound states of particles of different species. The unbroken density however corresponds to D3-D5 strings, which just translates into appropriate numbers of free particles in the field theory.

There may be some criticism over the name ``superconductivity'', as on the field theory side the $U(1)$ that is broken is only a global symmetry. 
First of all, the local structure is not relevant for the symmetry breaking pattern, but it is only important once one considers further perturbations or external fields. Also, there are many AdS/CFT papers that observed e.g. the delta function in the conductivity or the Meissner effect -- which are characteristic for local symmetry breaking. This should be also straightforward to implement here, as we could add a term like $A_{magn.}\, =\,B\, \tau_3\, x\, d\,y$ that will act like an external magnetic field and one can straightforwardly study the conductivity using linear response theory. For this, we would expect to obtain results similar to the ones in the literature.
%
%
%
%
\subsection{Perturbative Results}
First, in section \ref{pert} we considered an expansion of the non-abelian DBI action to the level $\alpha'^2$ in order to give us a quadratic action (with quartic coupling) of the relevant fields. At this level, there is actually no coupling in between the different condensates such that our approach of considering them separately is exact, and one could in principle construct a more complex phase diagram involving combinations of the different condensates though linear combination. 
Comparing all the energy gaps in section \ref{apert}, we found that the different phases look very similar, up to a scaling of the temperature dependence, and they all flow to the same $T=0$ diverging behavior with no crossing of phases. This might be due on the one hand to some supersymmetry effect, on the other hand, as we saw a $\Delta F\propto 1/T$ divergence, this is more likely just an artifact of the incompleteness and irregularity of the $\alpha'^2$ expansion at large fields. All these symmetry breaking phases clearly showed a ``superconducting'' second order phase transition. We also looked at the grand canonical ensemble in app. \ref{gcapppert}, where we found that the phases with broken symmetry would be actually disfavored if we had rather fixed the chemical potential.
All of the phases had various higher mode solutions, and for the ``compact'' s-wave, there were also two different boundary conditions that we could impose, depending on what we consider as the (vanishing) source and what as the (spontaneously appearing) condensate. The higher modes always had smaller critical temperature and a smaller energy gap.

The condensate with the largest critical temperature and largest energy gap was the compact scalar in which the condensate corresponds to some asymptotic $\tau_1$-valued ``separation'' of the probe branes from each other and from the D3 branes, discussed in sec. \ref{cpert}. This is similar to an effective mass that is generated for the fundamental fields but not necessarily for their bound states. This particular condensate (only the 0-mode) had a diverging value of the condensate as $T\rightarrow 0$, so the $T=0$ values of the condensates of the different modes were approximately proportional to $n^{-1/2}$. The condensate dual to the other boundary condition had a much lower critical temperature and showed an interesting flow of the value of the condensate to the same $T=0$ value for all the different modes. 

The p-wave condensate of sec. \ref{ppert} had the second highest critical temperature, but we did not discuss it in detail as it has been readily studied in the literature \cite{johannascon5}.
The s-wave phase corresponding to the separation in the ``flat'' directions that we discussed in section \ref{fpert} had the lowest critical temperature.
While the ``internal parameters'', the horizon value of the ``isospin electric flux'', i.e. the uncondensated part of the density, and the horizon value of the symmetry breaking field had a logarithmic dependence at small temperatures in the other cases, the scalar of the flat direction was special. There, we found an interesting attractor behavior to a finite fixed point. The condensate value of this scalar remains finite as $T\rightarrow 0$ -- computing it was numerically non-trivial due to the high order of the asymptotic scaling.
\subsection{Non-Perturbative Results}
In section \ref{fullact}, we considered the full non-abelian DBI action in the so-called adapted symmetrized trace prescription. We showed how this can be motivated from the symmetrization procedure, and it can also be motivated as it reflects the regularity of the DBI action and the geometry of the probe brane embedding. As such, it gives physically sensible results even though the details of the quantitative results may not be exact and there may be some other artifacts in special cases. Also for the unbroken phase this procedure is exact, as there is only one non-trivial generator in the DBI action.

Overall, we found in \ref{afullact}, that the $T\rightarrow 0$ divergences of the $\alpha'^2$ expansion, e.g. in the free energy, are cured from the full DBI action. Inspecting the entropy, it turns out that the residual $T=0$ entropy of the unbroken phase vanishes in any of the broken phases, i.e. the ground state degeneracy gets lifted. Another effect of the regularity of the action is also that the critical temperature of the smaller-$T_c$ condensates gets lowered further, such they the critical temperatures are approximately exponentially separated. However the ordering of the critical temperatures does not change.

While the p-wave phase that we discussed in sec. \ref{pfullact} does not change in its qualitative shape, the s-wave condensates do. Firstly, in all but the largest-$T_c$ phases, the second order phase transition changes to a first order one, and one of the compact s-waves crosses with the $n=0$ p-wave phase, and potentially the same happens at the higher modes. Also, the s-wave phases looked similar to each other and showed some multivaluedness that is usual for Minkowski - black hole (``meson melting'') phase transitions. In the ``flat'' scalar in section \ref{ffullact}, this was however a surprise. Still, we saw there the same, albeit now heavily distorted, attractor behavior of the internal parameters as in the $\alpha'^2$ expansion. 

For the compact scalar, we found that the black hole embeddings are dis-favored compared to the unbroken phase for all but the largest $T_c$ cases. Their embeddings reach eventually a critical point at which the flux through the horizon to the probe brane (the un-condensated density) vanishes, and after which they do not extend to the horizon. In those ``Minkowski'' embeddings, the isospin flux/density is almost entirely generated by the condensate (the D5-D5 strings) in addition to some spread-out D3-D5 strings -- in contrast to the abelian flux, where such embeddings would not be allowed \cite{robdens,fancytherm}. We found a remnant of the usual phase transition between those phases, but physically only a first order phase transition from the unbroken phase to the Minkowski phase takes place.
 
For the largest-$T_c$ phase there exists still the second order phase transition to the black hole phase. But rather than seeing the expected first order phase transition to the Minkowski phase that would extend to $T=0$, the gap in the free energy of the Minkowski phase and the value of the condensate diverge at a finite temperature. As shown in app. \ref{gcappfull}, at this point also the chemical potential vanishes and in the grand canonical ensemble there would be a broken phase in the UV limit.
This unusual divergence is perfectly well understood from the gravity side as we discuss in section \ref{combfullact} and it is purely due to the fact that only a finite isospin flux is required to support these embeddings. Hence it is likely to be an artifact for the following reason: If we consider the symmetrization of a term that contains only one generator, e.g. $\tau_1\tau_1\tau_1\tau_1$, the adapted procedure is exact for this term. Hence, the overall ``saturation'' effects will persist if one takes the full DBI action. For very much mixed terms, like $\tau_1\tau_2\tau_1\tau_2$, however there will be some cancellation due to the symmetrization, e.g. of $\tau_1\tau_2\tau_1\tau_2$ and $\tau_2\tau_1\tau_1\tau_2$, and the result of the symmetrized trace will be lower than the ``adapted'' version. This implies that we over-estimate the coupling of the symmetry breaking field to the isospin potential in particular in the region where the fields are of the same order. This applies in particular to this case where we find the divergence. As this happens only when there is a particular balance of the coupling term, we can expect that with a smaller coupling, the diverging phase would extend to large dimensionless density, i.e. vanishing temperature, where the divergence gets canceled by the overall density and temperature scaling, just as in the other phases.
%
%
\subsection{Outlook}
For future research, there may be three interesting avenues to take: Firstly, one could obviously consider more complicated configurations with more simultaneous fields. 
One particularly interesting one is the case where we explicitly separate the probe branes by turning on the $\tau_3$-valued scalar in the flat directions. This corresponds in the field theory to separating the different layers. Hence it is a very physical problem to address. It does not affect the unbroken phase, but in the broken phases, a quick look at the DBI action suggests that it will lower the critical temperatures. Most interestingly, it will affect the p-wave and s-wave condensates differently. In particular in the simplest setup it would couple to all the fields except for the scalar in the flat directions.

Another interesting thing to do would be to try to do a resummation of the expansion of the square root of the non-abelian DBI action after taking the symmetrized trace. In principle, this is unlikely to be successful, but this $SU(2)$ case may be special, as we can use the straightforward anticommutators of the Pauli matrices.

\acknowledgments The authors would like to thank
Johanna Erdmenger, Javier Tarrio, Martin Ammon, Rene Meyer, Petr Jizba, Sang-Jin Sin, Yumi Ko, Yunseok Seo and Xiaojian Bai
for useful comments and discussions,
the Erwin Schr\"odinger Institute for hospitality during the program on ``AdS Holography and the Quark-Gluon Plasma''
and the APCTP for hospitality during the APCTP-WCU Focus Program on From dense matter to compact stars in QCD and in hQCD.
This work was supported by the National Research Foundation of Korea(NRF) grant funded by the Korea government(MEST)
through the Center for Quantum Spacetime(CQUeST) of Sogang University with grant number 2005-0049409.
\appendix
\section{Grand Canonical Ensemble}\label{gcapp}
To give one more motivation for the choice of the canonical ensemble, for completeness and because it may be interesting in some contexts, we show some results related to the grand canonical ensemble.
\subsection{Perturbative Case}\label{gcapppert}
\myfigured{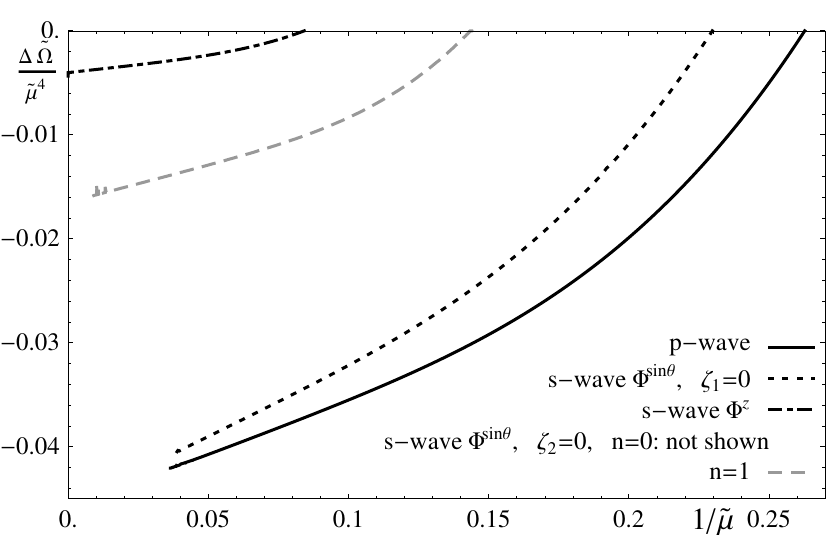}{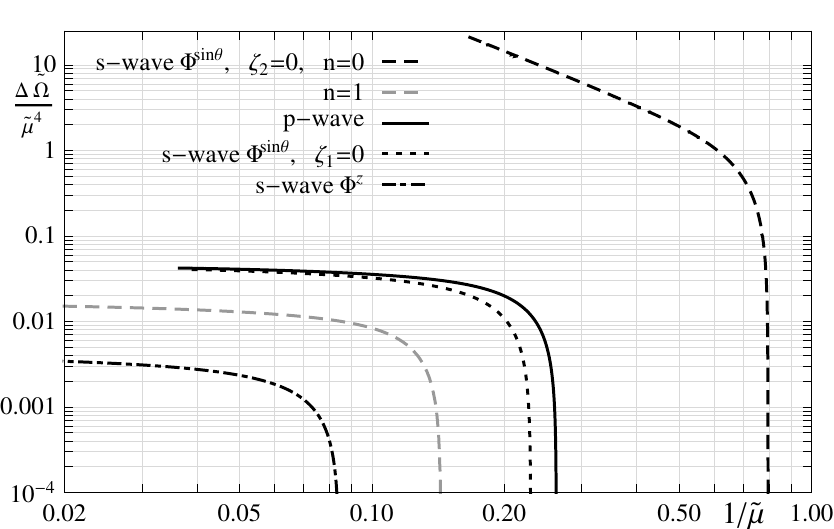}{The gap in the Gibbs free energy for the different condensates. As the gap is negative, we show the absolute value in the double logarithmic plot (right) and because of the strong divergence and large critical temperature of the $\zeta_2=0$, $n=0$ mode, we do not show this mode on the linear plot (left).}{compall_mu}
We show the Gibbs free energy as a function of the ``grand canonical temperature'' $1/\tlmu$ in fig. \ref{compall_mu}.
There we see that the behavior of the $\zeta_2=0$, $n=0$ mode is different from the other modes. Further, the energy gap is negative, so no symmetry breaking phase transition takes place and we see also no attractor behavior at small temperatures. Also, the first derivative is not continuous such that if there were a phase transition it would be first order. Still, this seems to be unphysical as the charge density in a superconductor (as opposed to e.g. a semiconductor) should be fixed. 
\subsection{Non-Perturbative Case}\label{gcappfull}
\myfigured{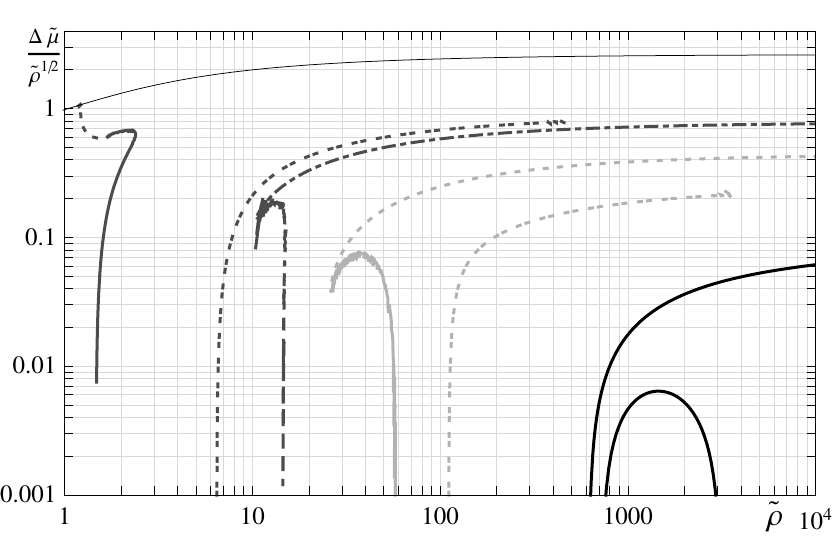}{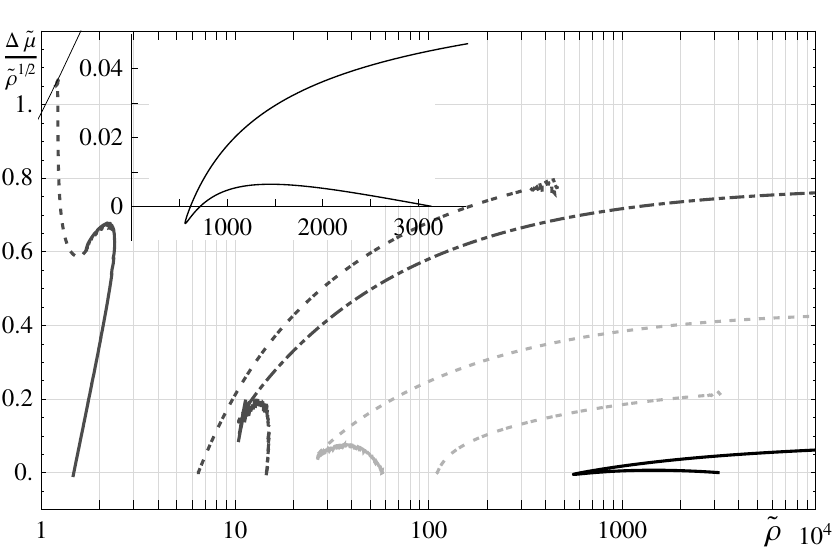}{Gap in the chemical potential $\Delta \mu = \mu_{unbroken}- \mu_{broken}$ as a function of density at fixed temperature. The narrow black curve shows the chemical potential of the unbroken phase; the other lines are the same as in the plot for the free energy. Left: Double-logarithmic plot. Right: Logarithmic-linear plot. The inset shows the details of the $\zeta_\psi$ s-wave phase.}{compfullmu}
Let us first look at the gap in the chemical potential $\Delta \mu = \mu_{unbroken}- \mu_{broken}$ as a function of density at fixed temperature. in fig. \ref{compfullmu}. There are three interesting observations: Firstly, in the limit of large densities, we see that $\Delta \tlmu$ is proportional to $\sqrt{\tlrho}$, such that the chemical potential is lowed by an approximately constant fraction. Secondly, the gap near the ``turning point'' of the s-wave phases decreases with decreasing critical temperature (increasing critical $\tlrho$), and in the phase of $\Phi^z$ $\Delta \tlmu$ is negative, i.e. the chemical potential increases, over a small range of density. This may be a feature of the regularity of the DBI action and corresponds also to the observation that the second order phase transitions turned into first oder ones. 

Thirdly, we see that the chemical potential actually vanishes at the ``endpoint'' of the $\varphi_2 = 0$, $n=0$ phase. This can be perfectly well understood from 
the geometry, as the charge density $\tlrho = - \lim_{u\rightarrow 0} \partial_u \rho(u)$ remains finite, but the maximum $u_0$ becomes small (the minimum radius
of the probe brane becomes large) and hence $\tlmu =\lim_{u\rightarrow 0} \rho(u) = \int_{u_0}^{0}\partial_u \rho(u)$ vanishes.
\myfigurer{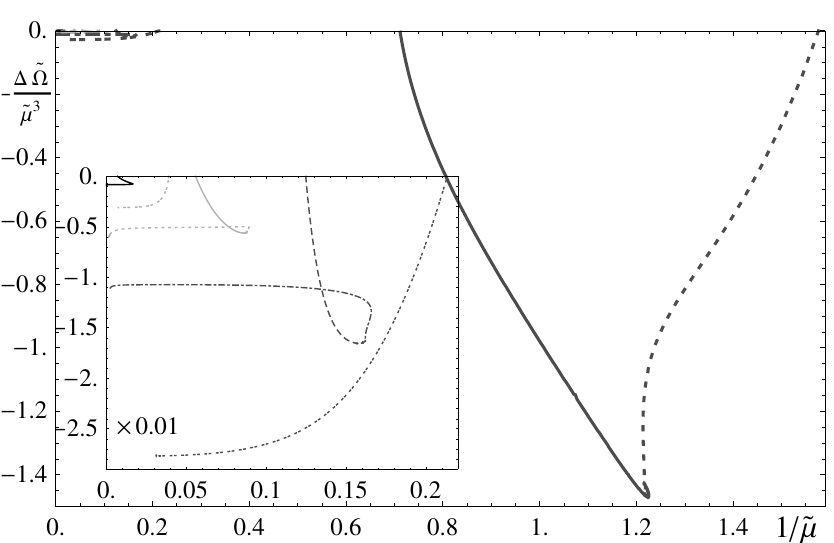}{The gap in the Gibbs free energy of the different condensates as a function of the ``grand canonical temperature'' $1/\tlrho$. The lines are the same as in the plot for the Helmholtz free energy, and the inset shows the details of the small-temperature phases.}{compfullom}

This is reflected in the plot of the \ref{compfullom} where we show the Gibbs free energy as a function of $1/\tlrho$, that serves as a temperature at fixed chemical potential. There, we see that the $\varphi_2 = 0$, $n=0$ phase extends out to infinite temperatures and $\delta \tilde{\Omega} = \tilde{\Omega}_{unbroken} - \tilde{\Omega}_{broken}$ even becomes positive above some critical temperature. This feature seems to be an unphysical artifact of the adapted symmetrized trace prescription and should disappear if one takes the rigorous symmetrized trace and possible higher order corrections, as we will discuss in the conclusions.

Else than that, the energy gap is negative in all the phases, so none of them is realized at fixed chemical potential and we see the lowering of the ``critical temperatures'' compared to the $\alpha'^2$ expansion. Further, all the s-wave phases change in a similar fashion to a shape similar to the plots of the condensates -- in the case of the compact scalar with a point at which the Minkowski and black hole phases intersect, similar to the meson melting phase transition \cite{long,fancytherm} in the abelian case.
\bibliography{mybibfile}
\bibliographystyle{JHEP}
\end{document}